\documentclass{ws-ijmpd}
\usepackage{epsfig}
\usepackage[spanish,english]{babel}
\usepackage{mathrsfs}

\newcommand{\E}{\mathcal{E}}
\newcommand{\Ham}{\mathcal{H}}

\newcommand{\lp}{\left(}
\newcommand{\rp}{\right)}

\begin{document}

\markboth{Gonzalo J. Olmo}
{Palatini Approach to Modified Gravity}

%
\catchline{}{}{}{}{}
%

\title{Palatini Approach to Modified Gravity: $f(R)$ Theories and Beyond}

\author{Gonzalo J. Olmo}

\address{Departamento de F\'{i}sica Te\'{o}rica and IFIC, Centro Mixto Universidad de Valencia - CSIC.
  Facultad de F\'{i}sica, Universidad de Valencia, Burjassot-46100, Valencia, Spain\\
gonzalo.olmo@uv.es}

\maketitle

\begin{history}
\received{}
\revised{}
\comby{Managing Editor}
\end{history}

\begin{abstract}
We review the recent literature on modified theories of gravity in the Palatini approach. After discussing the motivations that lead to consider alternatives to Einstein's theory and to treat the metric and the connection as independent objects, we review several topics that have been recently studied within this framework. In particular, we provide an in-depth analysis of the cosmic speedup problem, laboratory and solar systems tests, the structure of stellar objects, the Cauchy problem, and bouncing cosmologies. We also discuss the importance of going beyond the $f(R)$ models to capture other phenomenological aspects related with dark matter/energy and quantum gravity.
\end{abstract}


\section{Introduction}	
Einstein's theory of general relativity (GR) represents one of the most impressive exercises of human intellect. It implies a huge conceptual jump with respect to Newtonian gravity. The idea of gravitation as a force acting in an absolute space is replaced by a geometrical theory of space and time in which the space-time itself is a dynamical entity in interaction with the particles and fields living in it. This interaction is prescribed by a minimal coupling of those fields to the space-time metric according to what is today known as the Einstein equivalence principle (EEP). The dynamical equations for the gravitational field itself were deduced on grounds of mathematical simplicity and demanding that certain conservation laws were satisfied. Unlike the currently established standard model of elementary particles, no experiments were carried out to probe the structure of the theory. In spite of that, to date the theory has successfully passed all precision experimental tests. Its predictions are in agreement with experiments in scales that range from millimeters to astronomical units, scales in which weak and strong field phenomena can be observed\cite{Will-LR}. The theory is so successful in those regimes and scales that it is generally accepted that it should also work at larger and shorter scales, and at weaker and stronger regimes. \\

This extrapolation is, however, forcing us today to draw a picture of the universe that is not yet supported by other independent observations. For instance, to explain the rotation curves of spiral galaxies, we must accept the existence of vast amounts of unseen matter surrounding those galaxies. Additionally, to explain the luminosity-distance relation of distant type Ia supernovae and some properties of the  distribution of matter and radiation at large scales, we must accept the existence of yet another source of energy with repulsive gravitational properties\cite{reviews}. Together those unseen (or dark) sources of matter and energy are found to make up to $96\%$ of the total energy of the observable universe! This huge discrepancy between the gravitationally estimated amounts of matter and energy and the direct measurements via electromagnetic radiation motivates the search for alternative theories of gravity which can account for the large scale dynamics and structure without the need for dark matter and/or dark energy. In this sense, we note that the Newtonian accelerations felt by stars and gas clouds in the disk of spiral galaxies are orders of magnitude smaller than the accelerations measurable in laboratory. Thus there is no experimental evidence supporting the validity of Newton's law down to such tiny scales. 
For this reason, it seems legitimate and well justified to explore modifications of Newton's law and Einstein's theory to see if they can provide a consistent alternative picture of the observed Universe. In this directions we find Milgrom's proposal of Modified Newtonian dynamics\cite{MOND} (MOND), which fits surprisingly well many observational data in its natural regime of applicability and may be related with the Palatini approach\cite{Milgrom:2009ee}, and a variety of fully relativistic theories which can be used to do cosmology, such as $f(R)$ theories\cite{f(R)}, scalar-tensor theories, scalar-tensor-vector theories\cite{Bruneton:2007si}, higher-dimensional and brane-world scenarios\cite{branes}, \ldots \\

The extrapolation of the dynamics of GR to the very strong field regime indicates that the Universe began at a singularity and that the death of a sufficiently massive star unavoidably leads to the formation of a black hole singularity. Space-time singularities signal the breakdown of the theory, because the absence of a well-defined geometry implies the absence of physical laws and lack of predictability\cite{Novello-2008,Hawking-1975}. For this reason, it is generally accepted that the dynamics of GR must be changed at some point to avoid these problems. A widespread belief is that at sufficiently high energies the gravitational field must exhibit quantum properties that alter the dynamics and prevent the formation of singularities. In this sense, a perturbative approach to quantum gravity indicates that the Einstein-Hilbert Lagrangian must be supplemented by quadratic curvature terms to render the theory renormalizable\cite{Stelle-1977,Parker-Toms}. More recent approaches to quantum gravity, such as string theory, also regard GR as the low energy limit of a theory that should pick up increasing corrective terms at higher and higher energies\cite{strings}. The canonical quantization of GR using the so-called Ashtekar-Barbero variables\cite{LQG} predicts that the continuum space-time of GR is replaced by a quantum geometry in which areas and volumes are quantized in bits of an elementary unit of order the Planck scale. The low energy limit of this theory should also recover the classical dynamics of GR with corrections signaling the discreteness of the space-time. \\

The above discussion shows that there are theoretical and phenomenological reasons to explore the dynamics of alternative theories of gravity. Though dark matter and dark energy could play in cosmology a role similar to that played by the neutrino in the process of radioactive beta decay\footnote{Since conservation of energy and momentum was a pillar of special relativity, rather than proposing a modification of this principle, Pauli postulated the existence of a massless particle, the neutrino, to explain the spectrum of energies in the process of beta decay. Though in that case a ``dark matter'' particle solved the problem, to explain the anomalous perihelion shift of Mercury Einstein had to modify Newton's theory of gravity.}, we must try to figure out if our theory of gravity can be suitably corrected to explain the dynamics at large scales. Since we have well grounded reasons to believe that gravity must be modified in the ultraviolet regime, we should not be surprised by having to add corrections also at some infrared scale.

\subsection{The Palatini approach to modified gravity}

Because there are no limits to imagination, one should use experiments as a guide to constrain the range of possibilities to build an alternative theory of gravity. In this sense, the experimental efforts carried out in the 1960's to understand the nature of gravitation\cite{Dicke-1964} and the kind and properties of the fields associated to gravity, left it clear that gravitation is a geometric phenomenon. This led to the conclusion that the matter and the other non-gravitational fields must couple only to the metric, which implies that the total action must be of the form
\begin{equation}\label{eq:MTG}
S=S_G[g_{\alpha\beta},\phi, A_\mu,\ldots]+S_m[g_{\alpha\beta},\psi] \ ,
\end{equation}
where $g_{\alpha\beta},\phi, A_\mu,\ldots$, represent the gravitational fields (which can be scalars, vectors, and tensors of different ranks) and $\psi$ represents collectively the matter fields. This defines a class of theories known as ``metric theories of gravity\cite{Will-1993}'' which, by construction, should satisfy the EEP. 


Perhaps motivated by the restrictions imposed by the EEP, alternative theories of gravity have traditionally focused mainly on pseudo-Riemannian geometry, thus forcing the affine connection to be metric compatible. However, metricity and affinity are {\it a priori} logically independent concepts\cite{Zanelli} and, therefore, there is no fundamental theoretical reason to constrain the connection to be metric compatible. 
In fact, since affinities are very simple and fundamental geometrical entities\cite{Kobayashi-Nomizu}, in applying Ockham's razor to the construction of alternative theories of gravity we should give them higher priority than to other types of tensorial fields. For this reason, in this work we will mainly focus on  modified theories of gravity in which metric and connection are regarded as independent fields. Note that the existence of a metric in the theory naturally endows space-time with a Riemannian connection, the Levi-Civita connection $L^\alpha_{\beta\gamma}$. Thus, accepting the existence of an additional connection $\Gamma^\alpha_{\beta\gamma}$ is equivalent to having an independent rank-three tensor field $B^\alpha_{\beta\gamma}=\Gamma^\alpha_{\beta\gamma}-L^\alpha_{\beta\gamma}$ in the action. \\
Though gravitational redshift experiments do not impose very tight constraints on the possible coupling of a non-metric connection to matter\cite{Will-LR}, for simplicity we will follow the guide provided by experiments and will assume that freely falling bodies do follow geodesics of the metric. Thus, rather than working in a purely metric-affine framework in which matter is allowed to couple to the independent connection, we will consider only this restricted version, which is known as Palatini\footnote{For a discussion of this terminology see Ref.\refcite{FFR-1982}} formalism. This way we stick ourselves to the class of metric theories of gravity introduced above, in which the matter action is only coupled to the metric (and perhaps to its derivatives via the Levi-Civita connection) and the gravitational sector is of the form $S_G[g_{\alpha\beta},\Gamma^\alpha_{\beta\gamma},\phi, A_\mu,\ldots]$. \\

The {\it Palatini method} to obtain the field equations of GR was introduced by Einstein himself in 1925\cite{FFR-1982}. Despite considering independent variations of the metric and the connection, the resulting equations in GR turn out to be equivalent to those obtained doing variations of the metric only (metric variational formalism). This is so because the equation for the connection simply establishes its compatibility with the metric. However, this is just an accident. For other Lagrangians, in general, the field equations in metric and Palatini formalisms are different, as we will see in detail later (see Refs.\refcite{Met-Vs-Pal,Burton-Mann,IKPP} for some studies on the relation between Palatini and metric formalisms). But the differences between metric and Palatini formalisms go beyond the field equations, and this can already be seen in the context of GR. In fact, since the Einstein-Hilbert action contains second-order derivatives of the metric, to have a well defined variational principle one must add a surface term proportional to the extrinsic curvature, which explicitly refers to an embedding of the space-time into some background metric. In the Einstein-Palatini action, however, there are no derivatives of the metric and we only find first-order derivatives of the connection. As a result, it is usually claimed that no surface terms are necessary. However, to have a consistent formulation, conserved Hamiltonians at infinity in asymptotically flat spacetimes, and to correctly reproduce the thermodynamical properties of black holes, it has been recently found that a certain surface term must be added to the action. This surface term does not refer to any background, but when there is a background available, the metric and Palatini descriptions match\cite{Corichi-2010,Ashtekar-2008}. This implies that the corresponding path integral formulations of these two theories may be quite different. It is also worth noting that the consideration of the Einstein-Palatini action instead of the Einstein-Hilbert one was crucial for the implementation of the non-perturbative canonical quantization of the theory using Ashtekar variables\cite{LQG-book}. Therefore, the Palatini approach must be seriously considered not only to explore new phenomenological extensions of GR aimed at explaining the large scale structure of the universe, but also as a potential way to make contact with quantum gravity phenomenology. \\

\subsection{Goal and structure.}

In this article we will review the recent literature on modified theories of gravity framed within the Palatini formalism. Most of it will deal with theories of the $f(R)$ type, in which the gravity Lagrangian is given by a function of the scalar curvature $R$, but we will also comment on scalar-tensor theories and extensions of the $f(R)$ family that include other curvature invariants such as $R_{\mu\nu}R^{\mu\nu}$.
Though the Palatini approach had been considered in the past in different contexts, the interest in Palatini $f(R)$ theories was boosted by the observation\cite{Vollick-2003} that some of the known problems of the model\cite{CDTT} $f(R)=R-\mu^4/R$, proposed in metric formalism to explain the cosmic speedup, could be avoided by considering its Palatini version. Since then numerous works have addressed  different aspects of Palatini theories including the late-time and early-time cosmology, solar system and laboratory tests, stellar structure, the Cauchy problem, black hole thermodynamics, \ldots We have tried to provide a comprehensive and careful review of the literature on those topics that have received more attention. However, since we may have missed some useful and important references, we encourage the reader to help us in this task letting us know about those works. By ``careful'' we mean that we have tried to be as precise as possible in stating who did what and when, avoiding diffuse lists of references, which should help the newcomer find its way through the growing literature. We acknowledge that this procedure has been difficult in many cases and also that our criterion may not have been the best one at some points. Comments and suggestions in this respect will also be very welcome. \\

The content has been organized as follows. In Sec.\ref{sec:II}  we provide a detailed derivation of the field equations of Palatini $f(R)$ theories and discuss their scalar-tensor representation. Since space is limited, for the field equations of other Palatini theories we refer to the corresponding literature. Then we split the chronological evolution of the literature in subjects: the cosmic speedup problem is reviewed in Sec.\ref{sec:speedup}, laboratory and solar system tests are discussed in Sec.\ref{sec:local}, some questions related with stellar structure and new results regarding the Cauchy problem are analyzed in Sec.\ref{sec:others}, and the relation of Palatini theories with quantum gravity phenomenology is discussed in Sec.\ref{sec:QG}. We end with a summary and future perspectives.

\section{Field equations for Palatini theories \label{sec:II}}

Since most of the recent literature on Palatini theories has focused on $f(R)$ theories, we present here a detailed derivation of the field equations for this case. For extensions to actions containing other curvature invariants and couplings to scalar fields, we will refer to the corresponding literature. In this section we also comment on the scalar-tensor representation of $f(R)$ theories, the conservation of energy and momentum of matter, and some literature on other Palatini theories. 

\subsection{$f(R)$ theories}
The action of Palatini $f(R)$ theories is as follows
\begin{equation}\label{eq:f(R)-action}
S=\frac{1}{2\kappa^2}\int d^4x \sqrt{-g}f(R)+S_m[g_{\mu\nu},\psi] \ ,
\end{equation}
where $S_m$ is the matter action, $\psi$ represents collectively the matter fields, $\kappa^2$ is a constant with suitable dimensions (if $f(R)=R$, then $\kappa^2=8\pi G$), $R\equiv g^{\mu\nu}R_{\mu\nu}$,  $R_{\mu\nu}\equiv{R^\rho}_{\mu\rho\nu}$, and ${R^\alpha}_{\beta\mu\nu}=\partial_\mu\Gamma_{\nu\beta}^\alpha-\partial_\nu\Gamma_{\mu\beta}^\alpha+\Gamma_{\mu\lambda}^\alpha\Gamma_{\nu\beta}^\lambda-\Gamma_{\nu\lambda}^\alpha\Gamma_{\mu\beta}^\lambda$ represents the components of the Riemann tensor, the field strength of the connection $\Gamma^\alpha_{\mu\beta}$. Note that since the connection is determined dynamically, we cannot assume any {\it a priori} symmetry in its lower indices. This means that in the variation of the action to obtain the field equations we must bear in mind that $\Gamma^\alpha_{\beta\gamma}\neq  \Gamma^\alpha_{\gamma\beta}$. We will assume a symmetric metric tensor $g_{\mu\nu}=g_{\nu\mu}$ (for theories with non-symmetric $g_{\mu\nu}$ see, for instance, Ref.~\refcite{NGT}). The variation of the action (\ref{eq:f(R)-action}) with respect to the metric and the connection can be expressed as
\begin{eqnarray}\label{eq:var1-f)}
\delta S&=&\frac{1}{2\kappa^2}\int d^4x \sqrt{-g}\left[\left(f_R R_{(\mu\nu)}-\frac{f}{2}g_{\mu\nu} \right)\delta g^{\mu\nu} + f_R g^{\mu\nu}\delta R_{\mu\nu}\right]+\delta S_m \ ,
\end{eqnarray}
where $f_R\equiv \partial f/\partial_R$, and $R_{(\mu\nu)}$ represents the symmetric part of $R_{\mu\nu}$. 
Straightforward manipulations show that $\delta R_{\mu\nu}$ can  be written as
\begin{equation}
\delta R_{\mu\nu}= \nabla_\lambda \left(\delta \Gamma^\lambda_{\nu\mu}\right)-\nabla_\nu \left(\delta\Gamma^\lambda_{\lambda\mu}\right)+2S^\lambda_{\rho\nu}\delta\Gamma^\rho_{\lambda\mu} \ ,
\end{equation}
where $2S^\lambda_{\rho\nu}\equiv \Gamma^\lambda_{\rho\nu}-\Gamma^\lambda_{\nu\rho}$ represents the torsion tensor, the antisymmetric part of the connection. The contribution of the $\delta R_{\mu\nu}$ term, $I=\int d^4 x\sqrt{-g}f_R g^{\mu\nu}\delta R_{\mu\nu}$, leads to the following expression
\begin{equation}\label{eq:var2-f}
I=\int d^4 x\left[\nabla_\lambda\left(\sqrt{-g}J^\lambda\right)+\delta\Gamma^\lambda_{\nu\mu}\left\{-\nabla_\lambda\left(\sqrt{-g}f_R g^{\mu\nu}\right)+\nabla_\rho\left(\sqrt{-g}f_R g^{\mu\rho}\right)+2\sqrt{-g}f_R g^{\mu\sigma}S^\nu_{\lambda\sigma}\right\}\right] \ ,
\end{equation}
where $J^\lambda\equiv f_R\left(g^{\mu\nu}\delta\Gamma^\lambda_{\mu\nu}-g^{\mu\lambda}\delta\Gamma^\sigma_{\sigma\mu}\right)$. 
Having in mind that\cite{MTW} $\nabla_\mu \sqrt{-g}=\partial_\mu \sqrt{-g}-\Gamma^\sigma_{\mu\sigma}\sqrt{-g}$, we find that
\begin{equation}\label{eq:nabla-g}
\nabla_\lambda\left(\sqrt{-g}J^\lambda\right)=\partial_\lambda\left(\sqrt{-g}J^\lambda\right)+\sqrt{-g}f_R\left[g^{\mu\nu}S^\sigma_{\sigma\lambda}-\delta^\nu_\lambda g^{\mu\rho}S^\sigma_{\sigma\rho}\right]\delta\Gamma^\lambda_{\nu\mu} \ .
\end{equation}
Inserting this result in (\ref{eq:var2-f}) and assuming that the surface term $\int d^4 x \partial_\lambda\left(\sqrt{-g}J^\lambda\right)$ vanishes at the boundaries, the field equations can finally be written as follows
\begin{equation}\label{eq:field-g}
 f_R R_{(\mu\nu)}-\frac{f}{2}g_{\mu\nu} = \kappa^2T_{\mu\nu}
\end{equation}
\begin{equation}
-\nabla_\lambda\left(\sqrt{-g}f_R g^{\mu\nu}\right)+\delta^\nu_\lambda\nabla_\rho\left(\sqrt{-g}f_R g^{\mu\rho}\right)+2\sqrt{-g}f_R\left( g^{\mu\nu}S^\sigma_{\sigma\lambda}-\delta^\nu_\lambda g^{\mu\rho}S^\sigma_{\sigma\rho}+g^{\mu\sigma}S^\nu_{\lambda\sigma}\right)=H_\lambda^{\nu\mu}\label{eq:field-G}
\end{equation}
where $T_{\mu\nu}\equiv -\frac{2}{\sqrt{-g}}\frac{\delta S_m}{\delta g^{\mu\nu}}$, and $H_\lambda^{\nu\mu}\equiv -\delta S_m/\delta \Gamma^\lambda_{\nu\mu}=0$ because we assume that the matter is not coupled to the connection. To proceed further, it is common in the literature to impose the torsionless condition $S^\nu_{\lambda\sigma}=0$, which eventually turns (\ref{eq:field-G}) into the simpler form
\begin{equation}\label{eq:field-G-simpler}
\nabla_\lambda\left(\sqrt{-g}f_R g^{\mu\nu}\right)=0 \ .
\end{equation} 
However, with a bit of extra effort, we will gain deeper insight into the role and properties of the torsion and will see that an expression analogous to (\ref{eq:field-G-simpler}) can be reached without imposing any restriction on the torsion tensor. The first step is to trace over $\nu$ and $\lambda$ in (\ref{eq:field-G}) to get $3\nabla_\rho\left(\sqrt{-g}f_R g^{\mu\rho}\right)=4\sqrt{-g}f_Rg^{\mu\rho}S^\sigma_{\sigma\rho}$. We then insert this result in (\ref{eq:field-G}) to obtain
\begin{equation}
-\nabla_\lambda\left(\sqrt{-g}f_R g^{\mu\nu}\right)+2\sqrt{-g}f_R\left( g^{\mu\nu}S^\sigma_{\sigma\lambda}-\frac{\delta^\nu_\lambda}{3} g^{\mu\rho}S^\sigma_{\sigma\rho}+g^{\mu\sigma}S^\nu_{\lambda\sigma}\right)=0\label{eq:field2-G}
\end{equation}
We now split the connection into its symmetric and antisymmetric parts, which we denote $C^\lambda_{\mu\nu}$ and $S^\lambda_{\mu\nu}$ respectively, and reexpress $\nabla_\lambda\left(\sqrt{-g}f_R g^{\mu\nu}\right)$ in the form
\begin{equation}
\nabla_\lambda\left(\sqrt{-g}f_R g^{\mu\nu}\right)=\nabla^{C}_\lambda\left(\sqrt{-g}f_R g^{\mu\nu}\right)+\sqrt{-g}f_R\left[g^{\mu\sigma}S^\nu_{\lambda\sigma}+g^{\nu\sigma}S^\mu_{\lambda\sigma}+g^{\mu\nu}S^\sigma_{\sigma\lambda}\right] \ ,
\end{equation}
where $\nabla^{C}_\lambda\left(\sqrt{-g}f_R g^{\mu\nu}\right)$ only depends on the symmetric part of the connection, which means that $\nabla^{C}_\lambda A_\mu=\partial_\lambda A_\mu-C_{\lambda\mu}^\rho A_\rho$. Inserting this result in (\ref{eq:field2-G}), we get
\begin{equation}
\nabla^{C}_\lambda\left(\sqrt{-g}f_R g^{\mu\nu}\right)=\sqrt{-g}f_R\left(g^{\mu\sigma}S^\nu_{\lambda\sigma}-g^{\nu\sigma}S^\mu_{\lambda\sigma} +g^{\mu\nu}S^\sigma_{\sigma\lambda}-\frac{2}{3}\delta^\nu_\lambda g^{\mu\rho}S^\sigma_{\sigma\rho}\right) \ . \label{eq:field3-G} 
\end{equation}
Adding and subtracting to this equation the same expression but changing the order of $\mu$ and $\nu$ we find the following relations
\begin{eqnarray}
\nabla^{C}_\lambda\left(\sqrt{-g}f_R g^{\mu\nu}\right)&=&\sqrt{-g}f_R\left( g^{\mu\nu}S^\sigma_{\sigma\lambda}-\frac{1}{3}\left(\delta^\nu_\lambda g^{\mu\rho}+\delta^\mu_\lambda g^{\nu\rho}\right) S^\sigma_{\sigma\rho}\right)  \label{eq:addition} \\
g^{\mu\sigma}S^\nu_{\lambda\sigma}-g^{\nu\sigma}S^\mu_{\lambda\sigma}&=&\frac{1}{3}\left(\delta^\nu_\lambda g^{\mu\rho}-\delta^\mu_\lambda g^{\nu\rho}\right) S^\sigma_{\sigma\rho} \ . \label{eq:subtraction} 
\end{eqnarray}
Written in this way\footnote{Note that Eqs. (\ref{eq:field-G}) and (\ref{eq:field-G-simpler}) represent sets of $64$ independent relations which are equivalent to the $40$ relations of (\ref{eq:addition}) plus the $24$ relations of (\ref{eq:subtraction}).}, it is clear that the symmetric part of the connection is coupled to the antisymmetric part (the torsion) via the contraction $S^\sigma_{\sigma\rho}$. This term is also sourcing the right hand side of the torsion equation (\ref{eq:subtraction}). This fact suggests a new step aimed at simplifying the structure of (\ref{eq:addition}) and (\ref{eq:subtraction}). Consider the new variables
\begin{equation}\label{eq:newG}
\tilde{\Gamma}^\lambda_{\mu\nu}=\Gamma^\lambda_{\mu\nu}+\alpha \delta^\lambda_\nu S^\sigma_{\sigma\mu} \ ,
\end{equation}
and take the parameter $\alpha=2/3$, which implies that $\tilde{S}^\lambda_{\mu\nu}\equiv \tilde{\Gamma}^\lambda_{[\mu\nu]}$ is such that $\tilde{S}^\sigma_{\sigma\nu}=0$. The symmetric and antisymmetric parts of the connection $\tilde{\Gamma}^\lambda_{\mu\nu}$ are related to those of  ${\Gamma}^\lambda_{\mu\nu}$ by
\begin{eqnarray}\label{eq:newC}
\tilde{C}^\lambda_{\mu\nu}&=&C^\lambda_{\mu\nu}+\frac{1}{3}\left(\delta^\lambda_\nu S^\sigma_{\sigma\mu}+\delta^\lambda_\mu S^\sigma_{\sigma\nu}\right) \\
\tilde{S}^\lambda_{\mu\nu}&=&S^\lambda_{\mu\nu}+\frac{1}{3}\left(\delta^\lambda_\nu S^\sigma_{\sigma\mu}-\delta^\lambda_\mu S^\sigma_{\sigma\nu}\right) \label{eq:newS}
\end{eqnarray}
Rewriting (\ref{eq:addition}) and (\ref{eq:subtraction}) using these new variables, we find
\begin{eqnarray}
\nabla^{\tilde{C}}_\lambda\left(\sqrt{-g}f_R g^{\mu\nu}\right)&=&0  \label{eq:additionMod} \\
g^{\mu\sigma}\tilde{S}^\nu_{\lambda\sigma}-g^{\nu\sigma}\tilde{S}^\mu_{\lambda\sigma}&=&0 \ . \label{eq:subtractionMod} 
\end{eqnarray}
Written in this form (\ref{eq:subtractionMod}) implies that  $\tilde{S}_{\beta\lambda\alpha}=\tilde{S}_{\alpha\lambda\beta}$, where $g_{\beta\nu}\tilde{S}^\nu_{\lambda\alpha}\equiv \tilde{S}_{\beta\lambda\alpha}$. Since the torsion is antisymmetric in the last two indices, the symmetry of the first and third indices automatically implies that $\tilde{S}_{\beta\lambda\alpha}=0 \Leftrightarrow \tilde{S}^\nu_{\lambda\alpha}=0$. Using this result in (\ref{eq:newS}) we find that
\begin{equation}\label{eq:torsion}
S^\lambda_{\mu\nu}=\frac{1}{3}\left(\delta^\lambda_\mu S^\sigma_{\sigma\nu}-\delta^\lambda_\nu S^\sigma_{\sigma\mu}\right) \ .
\end{equation}
This result indicates that the torsion is generated by a vector $A_\mu\equiv S^\sigma_{\sigma\mu}$, which has important consequences and will be useful to solve (\ref{eq:additionMod}). The fact that 
\begin{equation}\label{eq:Conn}
\Gamma^\alpha_{\mu\nu}=\tilde{C}^\alpha_{\mu\nu} - \frac{2}{3} A_{\mu} \delta^\alpha_\nu \ 
\end{equation}
implies that $R^\alpha_{\beta\mu\nu}(\Gamma)=R^\alpha_{\beta\mu\nu}(\tilde{C})-\frac{4}{3}\partial_{[\mu}A_{\nu]}\delta^\alpha_\beta$, from which we get $R_{\mu\nu}(\Gamma)\equiv R^\alpha_{\mu\alpha\nu}(\Gamma)=R_{\mu\nu}(\tilde{C})-\frac{4}{3}\partial_{[\mu}A_{\nu]}$. From this it follows that the symmetric part of the Ricci tensor that appears in (\ref{eq:field-g}) is insensitive to the torsion because  $R_{(\mu\nu)}(\Gamma)= R_{(\mu\nu)}(\tilde{C})$. Obviously, $R$ is also insensitive to the existence of this type of torsion, i.e., $R(\Gamma)=R(\tilde{C})$. This property is known as the projective invariance of the scalar curvature\cite{Met-Affine,Querella:1998ke,SL-2006}. We are now ready to solve for (\ref{eq:additionMod}) [and (\ref{eq:field-G-simpler})]. Though (\ref{eq:additionMod}) seems to involve up to second-order derivatives of the connection, it can be reinterpreted using the trace of (\ref{eq:field-g}),
\begin{equation}\label{eq:trace-f}
R f_R-2f=\kappa^2T \ .
\end{equation}
This equation implies that $R=R(\Gamma)=R(\tilde{C})$ can be solved algebraically in terms of $T$, thus leading to $R=R(T)$ and $f_R=f_R(T)$, which are functions of the matter and possibly of the metric but not of the independent connection. 
The solution of (\ref{eq:additionMod}) can thus be easily found by defining a new metric $h_{\mu\nu}\equiv f_R(T) g_{\mu\nu}$ in terms of which that equation becomes simply $\nabla^{\tilde{C}}_\lambda(\sqrt{-h}h^{\mu\nu})=0$, which is an algebraic equation linear in the connection that leads to\cite{MTW}   
\begin{equation}\label{eq:LC-h}
\tilde{C}^\alpha_{\mu\nu}=\frac{h^{\alpha\rho}}{2}\left(\partial_\mu h_{\rho\nu}+\partial_\nu h_{\rho\mu}-\partial_\rho h_{\mu\nu}\right) \ .
\end{equation}
This completes our analysis of the equations that determine the connection $\Gamma^\alpha_{\mu\nu}$. We have found that, in general, $\Gamma^\lambda_{\mu\nu}$ is made out of a symmetric part, $\tilde{C}^\lambda_{\mu\nu}$, plus a vector-like contribution $- \frac{2}{3}\delta^\lambda_\nu A_{\mu}$. This vector is responsible for the existence of torsion, $S^\lambda_{\mu\nu}=\frac{1}{3}\left(\delta^\lambda_\mu A_{\nu}-\delta^\lambda_\nu A_{\mu}\right)$, but it does not affect the metric field equations (\ref{eq:field-g}) [see also Eq.(\ref{eq:Gab-f}) below], which justifies the usual approach in the literature of setting it to zero from the beginning. From this analysis it follows that the four conditions $A_\mu\equiv S^\sigma_{\sigma\mu}=0$ are enough to force the total vanishing of the torsion. When matter is coupled to the connection, the constraint $S^\sigma_{\sigma\rho}=0$ has also been suggested as a way to avoid potential inconsistencies of the field equations due to the projective invariance of the scalar curvature in GR\cite{Met-Affine} and in $f(R)$ theories\cite{SL-2006} (see also Ref.\refcite{FETG}). \\   

A non-trivial choice for the torsion vector can be motivated by introducing the expression (\ref{eq:LC-h}) for $\tilde{C}^\lambda_{\mu\nu}$  in (\ref{eq:Conn}). Using the relation $h_{\mu\nu}\equiv f_R g_{\mu\nu}$ one finds
\begin{equation}\label{eq:Conn-2}
\Gamma^\alpha_{\mu\nu}=L^\alpha_{\mu\nu}+\frac{1}{2f_R}\left[\delta^\alpha_\mu\partial_\nu f_R-g_{\mu\nu}\partial^\alpha f_R\right]- \frac{2}{3}\delta^\alpha_\nu\left(A_{\mu}-\frac{3}{4f_R}\partial_\mu f_R  \right)  \ ,
\end{equation}
where we have denoted $L^\alpha_{\mu\nu}\equiv \frac{g^{\alpha\rho}}{2}\left(\partial_\mu g_{\rho\nu}+\partial_\nu g_{\rho\mu}-\partial_\rho g_{\mu\nu}\right)$. 
Since the dynamics is insensitive to the presence of the vector $\tilde{A}_\mu\equiv A_\mu-\frac{3}{4f_R}\partial_\mu f_R$, one may wish to set $\tilde{A}_\mu=0$ to simplify the form of (\ref{eq:Conn-2}). By doing this, one finds that ${S^\alpha}_{\mu\nu}=\frac{\partial_\lambda f_R}{4f_R}(\delta^\alpha_\mu\delta^\lambda_\nu-\delta^\alpha_\nu\delta^\lambda_\mu)$ and  ${\Gamma^\alpha}_{\mu\nu}={L^\alpha}_{\mu\nu}+{K^\alpha}_{\mu\nu}$, where ${K^\alpha}_{\mu\nu}={S^\alpha}_{\mu\nu}+{{S_\mu}^\alpha}_\nu+{{S_\nu}^\alpha}_\mu$ is the so-called {\it contorsion} tensor. One can easily check [using for instance Eq. (\ref{eq:field2-G})] that this connection turns out to be compatible with the metric $g_{\mu\nu}$, i.e., it verifies $\nabla_\mu \left(\sqrt{-g}g^{\mu\nu}\right)=0$. This result shows that a torsionless $f(R)$ Palatini theory is dynamically equivalent to a metric-compatible $f(R)$ theory with torsion\cite{Capozziello:2007tj,Sotiriou:2009xt,CV-2009}. At the same time, those two particular cases are dynamically equivalent to a non metric-compatible Palatini $f(R)$ theory with arbitrary torsion generated by a vector field, which is the general case discussed here. \\

Now that we have an expression, Eq.(\ref{eq:Conn}), for the connection in terms of the metric, the matter, and the vector $A_\mu$, we can insert this solution for $\Gamma^\alpha_{\mu\nu}$ in (\ref{eq:field-g}) to obtain an equation that only involves the metric $g_{\mu\nu}$ and the matter: 
\begin{eqnarray}\label{eq:Gab-f}
R_{\mu \nu }(g)-\frac{1}{2}g_{\mu \nu }R(g)&=&\frac{\kappa
^2}{f_R}T_{\mu \nu }-\frac{Rf_R-f}{2f_R}g_{\mu \nu
}-\frac{3}{2(f_R)^2}\left[\partial_\mu f_R\partial_\nu
f_R-\frac{1}{2}g_{\mu \nu }(\partial f_R)^2\right]+ \nonumber \\ & &\frac{1}{f_R}\left[\nabla_\mu \nabla_\nu f_R-g_{\mu \nu }\Box
f_R\right] 
\end{eqnarray}
where $R_{\mu \nu }(g)$, $R(g)$, and $\nabla_\mu \nabla_\nu f_R$ are computed in terms of the
Levi-Civita connection of the metric $g_{\mu \nu }$, whereas $R$ and $f_R$ must be seen as functions of $T$.
To make our notation clearer, since $h_{\mu \nu }$ and $g_{\mu \nu }$ are
conformally related, it follows that $R=R(T)\equiv g^{\mu \nu }R_{\mu \nu
}(\Gamma)$ and $R(g)\equiv g^{\mu \nu }R_{\mu \nu }(g)$ are related by
\begin{equation}
R=R(g)+\frac{3}{2f_R}\partial_\lambda f_R\partial^\lambda
f_R-\frac{3}{f_R}\Box f_R
\end{equation}
where, recall, $f_R=f_R(T)$ is a function of $T$.  It is important to note that in vacuum, $T_{\mu\nu}=0$, the solution of (\ref{eq:trace-f}) is just a constant\footnote{Equation (\ref{eq:trace-f}) could have more than one solution, which could be interpreted as corresponding to different realizations of the Universe \cite{Mauro}. For simplicity, we assume that there exists only one physical solution, though one should bear in mind that particular models could have various solutions that were in agreement with observations in a certain regime.} $R_{vac}\equiv R(0)$, which implies that $f_R(0)$ is also a constant. As a consequence, the derivative terms on the right hand side of (\ref{eq:Gab-f}) vanish and that equation boils down to $G_{\mu\nu}=-\Lambda_{eff}g_{\mu\nu}$, where $\Lambda_{eff}\equiv\left.\frac{Rf_R-f}{2f_R}\right|_{R=R_{vac}}$ plays the role of an effective cosmological constant. This means that the dynamics of Palatini $f(R)$ theories departs from that of GR with a cosmological constant only in regions that contain sources, where $\frac{Rf_R-f}{2f_R}$ is no longer constant and the $\partial f_R$ terms are not zero. Therefore, it naturally follows that outside of the sources the solutions take the same form as those of GR with a cosmological constant, Birkhoff's theorem holds\cite{Faraoni:2010rt}, there are only two propagating degrees of freedom\cite{AMA-2009}, and there are no instabilities\cite{SotPLB} of the kind found in the metric version of these theories\cite{Dolgov:2003px}. Note, however, that the  conditions that such solutions must satisfy at the boundary separating the sources from the vacuum region will not, in general, be the same as in GR because the interior dynamics is different. \\
  
For some purposes, it may be useful to express the Palatini field equations (\ref{eq:Gab-f}) using the auxiliary metric $h_{\mu\nu}$ instead of $g_{\mu\nu}$. Taking into account that $R=g^{\mu\nu}R_{\mu\nu}(\Gamma)$ is related with $R(h)\equiv h^{\mu\nu}R_{\mu\nu}(\Gamma)$ by $R(h)=R/f_R$, (\ref{eq:field-g}) can be put as
\begin{equation}\label{eq:Gab-fE}
G_{\mu\nu}(h)=\frac{\kappa^2}{f_R(T)}T_{\mu\nu}-\Lambda(T)h_{\mu\nu} \ ,
\end{equation}
where $\Lambda(T)\equiv (Rf_R-f)/2f_R^2=(f+\kappa^2T)/2f_R^2$. It is worth noting that the conformal transformation needed to go from the representation (\ref{eq:Gab-f}) (the so-called Jordan frame) to the representation (\ref{eq:Gab-fE}) (Einstein frame) has absorbed all the terms with derivatives that appeared on the right hand side of (\ref{eq:Gab-f}), which makes simpler the manipulations of the field equations (\ref{eq:Gab-fE}). If one decides to forget about the original physical motivations that led to construct the $f(R)$ theory in the Jordan frame and chooses to interpret the Einstein frame metric $h_{\mu\nu}$ as the physical metric that defines free particle geodesics (which implies a redefinition of physical observables) then (\ref{eq:Gab-fE}) can be seen as a theory with a density-dependent effective Newton's constant and a varying cosmological {\it constant} $\Lambda(T)$. This possibility has also received some attention in the literature\cite{Paplowski,CDLFM,CaDaVer,Shojai:2008zz,Allemandi:2004yx}. \\

A final comment regarding the vacuum equations is in order. It is easy to see that the connection (\ref{eq:LC-h}) is invariant under a constant rescaling of the metric $h_{\alpha\beta}\to \lambda h_{\alpha\beta}$ and that $G_{\mu\nu}(h_{\alpha\beta})=G_{\mu\nu}(\lambda h_{\alpha\beta})$. If we now compare equations (\ref{eq:Gab-f}) and (\ref{eq:Gab-fE}) in vacuum, we find that $G_{\mu\nu}(g)=-\Lambda_{eff} g_{\mu\nu}=-\tilde{\Lambda}_{eff} h_{\mu\nu}$, with $\Lambda_{eff}=f_R(0)\tilde{\Lambda}_{eff}$. For the discussion of local experiments and stellar structure, it turns out to be convenient to rescale the metric in such a way that $g_{\mu\nu}=h_{\mu\nu}$ in vacuum. This is simply achieved by taking $g_{\mu\nu}=\frac{f_R(0)}{f_R(T)}h_{\mu\nu}$. This leads to $\Lambda_{eff}=\tilde{\Lambda}_{eff}$, which simply states that both constants are measured in the same units. This simple observation makes it clear that the difference in the dynamics of Palatini $f(R)$ theories in Einstein and Jordan frames amounts to a matter-induced local rescaling of units, i.e., the units used in Einstein and Jordan frames differ by a factor that depends on the local energy-momentum density. With the constant rescaling, $g_{\mu\nu}=\frac{f_R(0)}{f_R(T)}h_{\mu\nu}=\frac{1}{\phi(T)}h_{\mu\nu}$, the field equation (\ref{eq:Gab-fE}) can be put as follows
\begin{equation}\label{eq:Gab-fEr} 
G_{\mu\nu}(h)=\frac{\tilde{\kappa}^2}{\phi(T)}T_{\mu\nu}-\tilde{\Lambda}(T)h_{\mu\nu} \ ,
\end{equation}
where $\tilde{\kappa}^2\equiv\kappa^2/f_R(0)$, and $\tilde{\Lambda}(T)\equiv(f/f_R(0)+\tilde{\kappa}^2T)/2\phi^2$. \\

\subsection{Scalar-tensor representation of $f(R)$ theories}

The equations of motion (\ref{eq:Gab-f}) derived above can be rewritten as those of a usual (metric-compatible and torsionless) Brans-Dicke scalar-tensor theory,
\begin{eqnarray} \label{eq:ST}
S[{g}_{\mu \nu},\phi,\psi_m]&=&\frac{1}{2\kappa ^2 }\int d^4x\sqrt{-{g}}\left[\phi {R}({g})-\frac{\omega}{\phi}(\partial_\mu \phi\partial^\mu\phi)-V(\phi)\right]+S_m[{g}_{\mu \nu},\psi_m] \ , 
\end{eqnarray}
by just introducing the following notational change
\begin{equation}\label{eq:ST-form}
\phi\equiv f_R \ , \ V(\phi)\equiv Rf_R-f
\end{equation}
where in order to express $V=V(\phi)$ we assume invertible\footnote{Note that, unlike other derivations of the scalar-tensor representation, our manipulations do not impose any constraint on $f_{RR}$. See, for instance, Ref.~\refcite{Olmo2007a} for further details on this.} the relation $\phi=f_R$ to obtain $R=R(\phi)$. 
The equations of motion (\ref{eq:Gab-f}) for the metric  then become
\begin{equation}\label{eq:Gab-ST}
G_{\mu \nu }(g)=
\frac{\kappa^2}{\phi}T_{\mu\nu}-\frac{1}{2\phi}g_{\mu\nu}V(\phi)+\frac{\omega}{\phi^2}\left[\partial_\mu\phi\partial_\nu\phi-\frac{1}{2}g_{\mu\nu}(\partial \phi)^2\right]+\frac{1}{\phi}\left[\nabla_\mu\nabla_\nu\phi-g_{\mu\nu}\Box \phi\right]
\end{equation}
where in our case the constant parameter $w$ takes the value $\omega=-3/2$. In the Brans-Dicke theory, the scalar field $\phi$ is governed by the following equation 
\begin{equation}\label{eq:phi-BD}
(3+2\omega)\Box \phi +2V(\phi)-\phi \frac{dV}{d\phi}=\kappa^2T \ , 
\end{equation}
which using $w=-3/2$ boils down to 
\begin{equation}\label{eq:phi-ST-Pal}
2V(\phi)-\phi \frac{dV}{d\phi}=\kappa^2T \ .
\end{equation}
 This equation is the same as (\ref{eq:trace-f}) but written using the notational change introduced in (\ref{eq:ST-form}). It is interesting to note that $f(R)$ theories in metric formalism also admit a Brans-Dicke-like representation\cite{TT} in which $w$ turns out to be $w=0$. To our knowledge,  the identification of Palatini $f(R)$ theories with the case $w=-3/2$ was first carried out in Ref.~\refcite{Wang:2004pq}, though a scalar-tensor representation was already known\cite{Flanagan-2004a,Flanagan-2003}. The extension of this result to the metric-compatible $f(R)$ case with torsion was first given in Ref.\refcite{Capozziello:2007tj}, and to the more general case discussed here in Ref.\refcite{Sotiriou:2009xt} (see also Refs.\refcite{CV-2009,FV-2010,Dadhich:2010xa} for related works). Though this scalar-tensor representation can be useful for some considerations like the computation\cite{Olmo2005} and discussion\cite{Sot06} of the Newtonian and post-Newtonian limits and black hole thermodynamics\cite{BH-thermo} it should not be taken beyond its natural context. In fact, though one may be tempted to interpret Palatini   
$f(R)$ as the limiting case $w\to-3/2$ of the Brans-Dicke theory\cite{IKPP}, the fact is that the theory corresponds
exactly to the precise value $w=-3/2$. For this reason, the absence of dynamics for the corresponding
scalar field (absence of the $\Box\phi$ term) is not an issue of fine tunning, and the relation between it and the matter
needs not necessarily be interpreted as a strong coupling regime in which matter is 
forced to satisfy certain constraints to avoid exciting the $\Box\phi$ term. If the scalar field equation is read in its original
$f(R)$ form, its meaning and implications are much more transparent. Equation (\ref{eq:trace-f}) means that geometrical
objects such as the scalar $R$ are algebraically related with the matter sources in a way that depends on the form of the Lagrangian $f(R)$. In GR this relation is linear, $R=-\kappa^2T$, but in other theories it may be non-linear, $R=R(T)$. As we will see, that relation may
end up imposing constraints on the geometry, which obviously may back-react conditioning the dynamics of the matter fields. This interpretation is naturally extended to more general Palatini theories which do not admit a scalar-tensor representation\cite{OSAT,BO-2010}.
Therefore, in the Palatini version of $f(R)$ theories, unlike in the metric formalism, the independent connection does not introduce new dynamical degrees of freedom. Rather, it modifies the way matter generates the space-time curvature associated with the metric by generating new matter terms on the right hand side of the field equations. \\  

\subsection{Conservation of energy and momentum}

In Palatini $f(R)$ theories, like in all metric theories of gravity of the form (\ref{eq:MTG}), the conservation of the energy-momentum tensor is naturally satisfied and follows from the invariance under diffeomorphisms of the matter action\cite{HB-1993,Dick-1993,BDG-1999}. This can be seen as follows. Consider the variation of the action under an infinitesimal change of coordinates $\delta x^\mu=\epsilon^\mu(x)$
\begin{equation}
\delta S_m=\frac{1}{2}\int d^4x \frac{\delta \left(\sqrt{-g}\mathcal{L}_m\right)}{\delta g_{\mu\nu}}\delta g_{\mu\nu} \ .
\end{equation}
Since the (canonical) energy momentum tensor is defined as $T^{\mu\nu}=\frac{2}{\sqrt{-g}} \frac{\delta \left(\sqrt{-g}\mathcal{L}_m\right)}{\delta g_{\mu\nu}}$, and a diffeomorphism induces a change in the metric of the form $\delta g_{\mu\nu}=2\nabla_{(\mu} \epsilon_{\nu)}$, where $\nabla_\mu$ is the usual derivative operator involving the Christoffel symbols of the metric $g_{\mu\nu}$, it follows that
\begin{equation}
\delta S_m=\frac{1}{2}\int d^4x \sqrt{-g}T^{\mu\nu}\nabla_\mu\epsilon_\nu \ .
\end{equation}
If the matter action is invariant under diffeomorphisms, $\delta S_m=0$, then an integration by parts leads to
\begin{equation}\label{eq:conservation}
\delta S_m=-\frac{1}{2}\int d^4x \nabla_\mu\left(\sqrt{-g}T^{\mu\nu}\right)\epsilon_\nu=0 \ .
\end{equation}
Since $\nabla_\mu\left(\sqrt{-g}T^{\mu\nu}\right)=\sqrt{-g}\nabla_\mu T^{\mu\nu}$ and $\delta S_m$ vanishes for arbitrary $\epsilon_\nu$, (\ref{eq:conservation}) implies that $\nabla_\mu T^{\mu\nu}=0$. Note that despite this elementary result, it has sometimes been claimed that Palatini $f(R)$ theories do not satisfy the conservation of energy-momentum\cite{KreAl}. The covariant conservation of energy-momentum in very general modified theories of gravity was studied in detail in Ref.\refcite{Koivisto:2005yk}, in which theories with non-minimal couplings\cite{non-minimal} between the matter Lagrangian and the curvature were also considered.

\subsection{Other Palatini theories. } 

In the introduction we emphasized the fundamental role that the connection should play in the construction of alternative theories of gravity. In this sense, it is remarkable that the consideration in the recent literature of $f(R)$ theories in metric formalism was naturally followed by their Palatini counterpart. However, even though they are equally justified, scalar-tensor  and higher-dimensional theories (to name a few) in the Palatini approach have not received the same attention, and the literature in these subjects is scarce. We just find some studies of the conditions in which metric and Palatini scalar-tensor theories lead to the same field equations\cite{Burton-Mann}, on how dimensional reduction in $5-$dimensional Kaluza-Klein theory compares with the   metric approach\cite{DMHS}, an attempt to unify gravitation and electromagnetism in a $5-$dimensional quadratic curvature model\cite{Baskal:2010sv}, and some applications to inflationary cosmology\cite{Bauer-2008} and its perturbations\cite{Tamanini:2010uq} in $f(R)$ and scalar-tensor models. \\

The Palatini approach has been recently used  by Milgrom\cite{Milgrom:2009ee} in the context of modified Newtonian dynamics (MOND), which could open new avenues for the phenomenology of Palatini  theories in the context of dark matter. Milgrom's recent approach consists on expressing the Lagrangian formulation of Newtonian gravity using a Palatini approach and then introducing the necessary modifications to implement the MOND equations. This is done considering in the Lagrangian density 
\begin{equation}\label{eq:MOND_L}
L_N=\frac{1}{\kappa^2}\left(\vec{g}^2-2\phi \vec{\nabla}\vec{g}\right)+\rho\left(\frac{1}{2}\vec{v}^2-\phi\right) 
\end{equation}  
where $\rho=\sum_i m_i\delta(\vec{x}-\vec{x}_i)$, independent variations with respect to the variables $\vec{g}$ and $\phi$. Variation over $\vec{g}$ yields $\vec{g}=-\vec{\nabla}\phi$, and over $\phi$ gives $\vec{\nabla}\vec{g}=-\kappa^2\rho/2$, which yields Newtonian dynamics. A MOND-like theory is obtained by introducing the acceleration scale $a_0\sim 10^{-10}$ m/s$^2$ and replacing $\vec{g}$ by $a_0^2Q(\vec{g}^2/a_0^2)$ in (\ref{eq:MOND_L}), where the function $Q(z)$ must be such that $Q(z)\to z+$constant for large $z$, to recover the standard Newtonian laws at high accelerations, and $Q(z)\approx (4/3)z^{3/4}$ for $z\ll 1$, to produce MOND at low accelerations. The resulting equations are
\begin{eqnarray}
\ddot{\vec{x}}_i&=&-\vec{\nabla}\phi(\vec{x}_i) \\
\vec{\nabla}\vec{g}&=&-\frac{\kappa^2}{2}\rho \\
\vec{\nabla}\phi &=&-\nu\left(\left|\frac{\vec{\nabla}\vec{g}}{a_0}\right|\right)\vec{g} \label{eq:g_MOND} \ ,
\end{eqnarray}
where $\nu(z)\equiv d Q(z)/d z$. The above equations indicate that particles move according to the standard Newtonian law of inertia in the potential $\phi$. However, as it follows from (\ref{eq:g_MOND}), the MOND acceleration field\footnote{Strictly speaking, to recover MOND one should impose the further constraint $\vec{g}=-\vec{\nabla}\phi_N$.}  $\vec{g}_{MOND}\equiv -\vec{\nabla}\phi$ turns out to be an {\it algebraic} function of the Newtonian acceleration field $\vec{g}$. It is worth noting, as pointed out in Ref.\refcite{Milgrom:2009ee}, that one can get the gravitational part of the Lagrangian (\ref{eq:MOND_L}) from the non-relativistic limit of the Palatini formulation of GR. In this sense, we want to remark that since connections play the role of gravitational accelerations, as is clearly seen from the geodesic equation $d u^\mu/d\tau+\Gamma^\mu_{\alpha\beta}u^\alpha u^\beta=0$, whereas the metric is related to the potential field via $g_{00}\approx -1+2\phi$, the failure to satisfy the standard Newtonian acceleration law could be seen as a manifestation of connection-related effects. For generalizations and relativistic extensions of the theory presented here, see Ref.\refcite{Milgrom:2009ee} (see also section \ref{sec:beyond} for some related results).

\section{Cosmic speedup in Palatini $f(R)$ theories \label{sec:speedup}}

Observations of the cosmic microwave background (CMB) radiation\cite{CMB}, high redshift supernovae surveys\cite{SNIa}, large scale structure\cite{LSS}, and baryon acoustic oscillations\cite{BAO} suggest that the expansion history of the universe has passed through a number of phases, which consist on an earlier stage of rapidly accelerated expansion (known as inflation) followed by two periods of decelerated expansion dominated by the presence of radiation and dust (matter without pressure), respectively, and a current phase of accelerated expansion that started some five billion years ago following the era of matter domination. The field equations of GR in a Friedmann-Robertson-Walker (FRW) spacetime with line element $ds^2=-dt^2+a^2d\vec{x}^2$ filled with non-interacting perfect fluids of density $\rho_i$ and pressure $P_i$, 
\begin{equation}
\left(\frac{\dot{a}}{a}\right)^2+\frac{K}{a^2}=\frac{\kappa^2}{3}\rho \ , \hspace{0.5cm} \ \frac{\ddot{a}}{a}=-\frac{\kappa^2}{6}(\rho+3P) \ ,
\end{equation}
where $K$ is the spatial curvature, $\rho=\sum_i \rho_i$, and $P=\sum_i P_i$, 
indicate that a phase of positive accelerated expansion can only happen if there exists some matter/energy source that dominates over the others and whose equation of state satisfies $P_X/\rho_X< -1/3$, where $P_X$ and $\rho_X$ represent the pressure and energy density of that source. A natural candidate to explain the current phase of cosmic acceleration is a cosmological constant $\Lambda$, for which $P_\Lambda/\rho_\Lambda=-1$. However, this simple proposal is hard to accept from a theoretical point of view. If $\Lambda$ represents a new fundamental constant of Nature, one could expect new physical phenomena at cosmic scales in analogy with what happened when the Planck constant was discovered. If it is seen as vacuum quantum energy, then it is generally claimed that its observed value is too small to be in agreement with a naive estimation from quantum field theory, though if we apply more rigorous techniques of quantum field renormalization in curved space-times the predicted value turns out to be much smaller\cite{Hollands-Wald} than the observed one. For these and other reasons, there seems to be a widespread desire to explain the current cosmic speedup by means of some dynamical entity rather than by a pure constant of cosmic nature. \\

\begin{figure}[ht]
\begin{tabular}{lr}
{\psfig{file=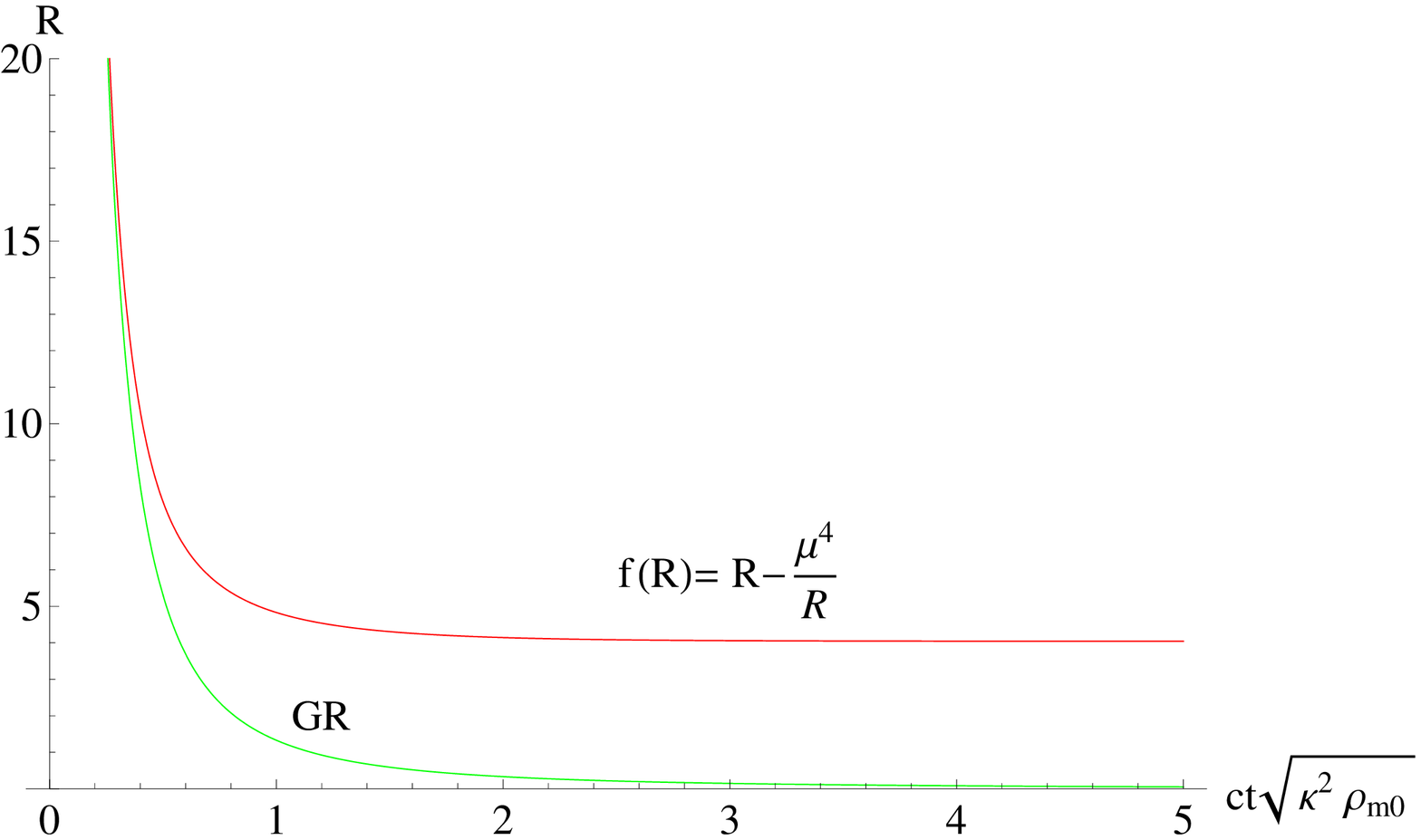,width=6.5cm}} & {\psfig{file=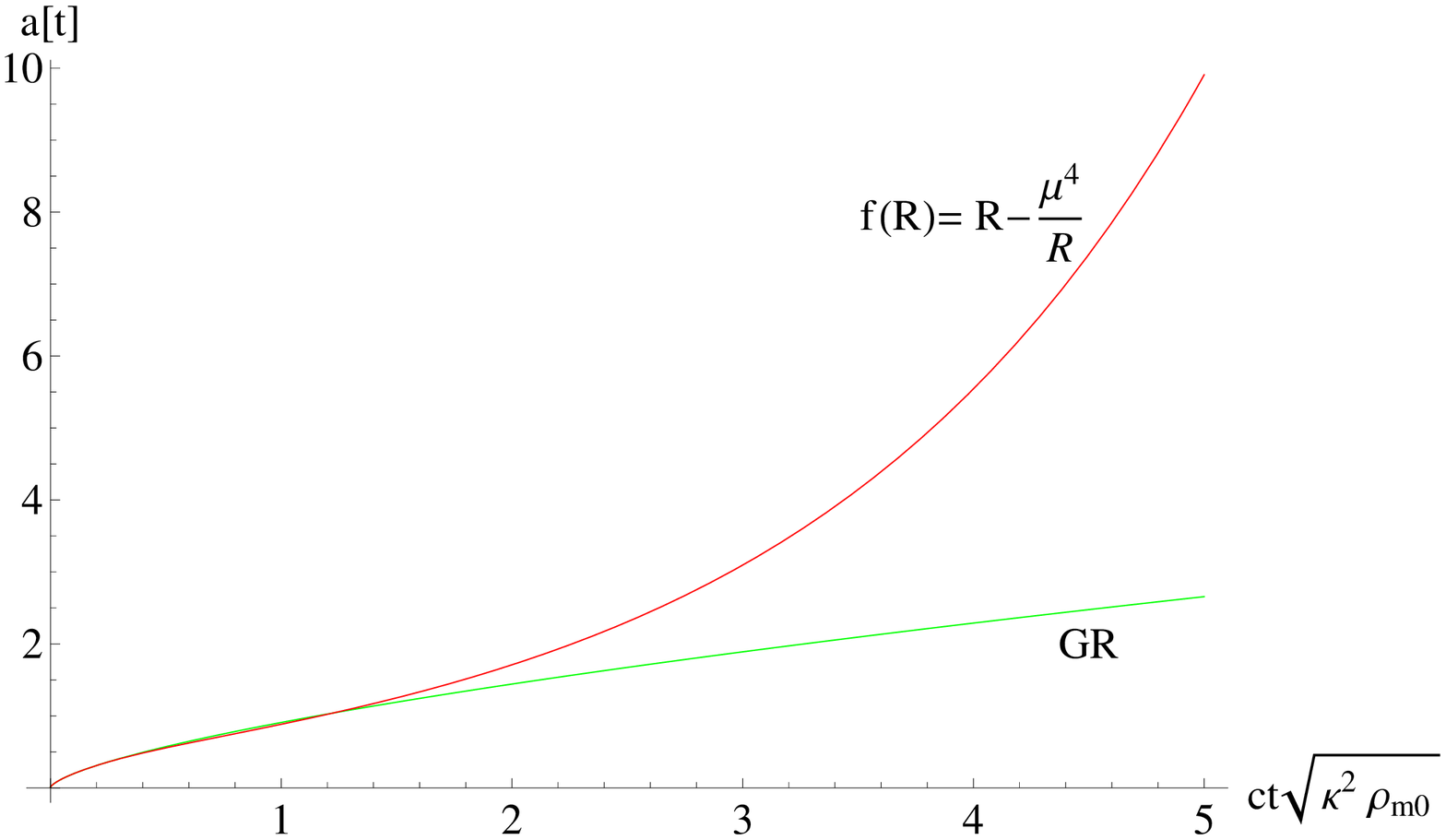,width=6.5cm}}
\end{tabular}
\vspace*{8pt}
\caption{Comparison of the time evolution of the curvature (left) and the expansion factor (right) in GR and the $1/R$ model with the initial condition $\left.a(t)\right|_{t=1}=1$. We took $\rho_\mu/\rho_{m_0}=0.7/0.3$. In the $1/R$ theory the curvature tends to a constant at late times, thus implying a de Sitter phase.\label{CDTT}}
\end{figure}

The fact that the field equations of Palatini $f(R)$ theories in vacuum exactly boil down to those of GR with an effective cosmological constant turned these theories into a very natural candidate\footnote{It should be noted that the Palatini dynamics is radically different from that corresponding to $f(R)$ theories in metric formalism. In that case, the modified dynamics is due to the existence of an additional effective scalar degree of freedom which is non-minimally coupled to the scalar curvature. This coupling turns the metric version of $f(R)$ theories into a particular type of extended quintessence model\cite{Perrotta:1999am} and, therefore, the metric $f(R)$ predictions are indistinguishable from that type of dark energy models.} to explain the cosmic speedup. For suitable choices of the function $f(R)$, it could happen that the new gravitationally-induced matter terms that appear on the right hand side of (\ref{eq:Gab-f}) were negligible during earlier phases of the expansion history but became dominant at later times, thus allowing an expansion that closely resembles GR in the past but produces cosmic speedup today. One could thus explain the transition from a matter dominated universe to an asymptotically de Sitter accelerated one with standard sources of matter and radiation but without the theoretical problems posed by a strictly constant $\Lambda$. The most famous $f(R)$ model of this kind investigated in the Palatini approach was borrowed from a proposal of Carroll et al.\cite{CDTT} in metric formalism, namely, $f(R)=R-\mu^4/R$, where $\rho_\mu\equiv\mu^2/\kappa^2$ represents the energy-density scale at which the effects of the modified dynamics are relevant. Vollick\cite{Vollick-2003} considered this model and showed that after the standard matter-dominated era, the expansion approaches a de Sitter phase exponentially fast. To see this, consider the modified Friedmann equation corresponding to a given $f(R)$ Lagrangian in a universe filled with matter and radiation
\begin{equation}\label{eq:Hubble-iso-f(R)}
H^2=\left(\frac{\dot{a}}{a}\right)^2=\frac{1}{6f_R}\frac{\left[f+\kappa^2(\rho_m+2\rho_r)-\frac{6K f_R}{a^2}\right]}{\left[1+\frac{3}{2}\frac{\kappa^2\rho_m f_{RR}}{f_R(Rf_{RR}-f_R)}\right]^2} \ ,
\end{equation}
where $\rho_m$ represents the energy density of the (pressureless) matter, $\rho_r$ is the energy density of radiation, and $R$ is a function of $\rho_m$ only because $T=-\rho_m$. In the $1/R$ model, one finds 
\begin{equation}\label{eq:R-CDTT}
R=\frac{\kappa^2\rho_m}{2}\left(1+\sqrt{1+12\left(\frac{\rho_\mu}{\rho_m}\right)^2}\right) \ , 
\end{equation}
which recovers $R\approx \kappa^2\rho_m$ when ${\rho_\mu}/{\rho_m}\ll 1$ and tends to the constant value $R_{vac}=\sqrt{3}\mu^2$ when ${\rho_\mu}/{\rho_m}\gg 1$ (see Fig.\ref{CDTT}). We thus see that when ${\rho_\mu}/{\rho_m}\ll 1$ then (\ref{eq:Hubble-iso-f(R)}) behaves as
$H^2\approx H^2_{GR}-\kappa^2(\rho_m+4\rho_r/3)(\rho_\mu/\rho_m)^2+\ldots$, which is virtually indistinguishable from GR. However, when the matter energy density, $\rho_m\sim a^{-3}$, drops below the constant value $\rho_\mu$, ${\rho_\mu}/{\rho_m}\gg 1$, then (\ref{eq:Hubble-iso-f(R)}) goes like $H^2\approx \frac{\mu^2}{4\sqrt{3}}+\frac{19}{96}\kappa^2\rho_m+\ldots$, which tends to a constant and implies an asymptotically de Sitter expansion, thus confirming the late time cosmic speedup (see Fig.\ref{CDTT}).  

\subsection{Cosmological constraints}

The $1/R$ model was soon compared with observations of type Ia supernovae\cite{Meng-Wang}, though such first studies, as we will see, were excessively optimistic about its viability. This optimism may have its origin in earlier studies of Palatini $f(R)$ cosmologies which concluded that these theories were very poorly constrained\cite{BHV-2002}, being $|f_{RR}(0)|<10^{113}$ one of the constraints coming from cosmological data. Besides the $R-\mu^4/R$ theory, which represented a small departure from GR at low matter densities, some authors also explored whether radical departures from the GR dynamics at cosmic scales such as $f(R)=\beta R^n$ or $f(R)=\alpha\ln R$ could be compatible with observations. These models were confronted with the Hubble diagram of type Ia Supernovae, the data on the gas mass fraction in relaxed galaxy clusters\cite{CCF-2004}, and baryon acoustic oscillations\cite{BGS-2006}. Though the fits to the data were good, the statistical analysis did not suggest any improvement with respect to the standard $\Lambda$CDM model. On the other hand, tight constraints on the family of models $R-\alpha R^\beta$ were obtained by studying the cosmic microwave background (CMB) shift parameter and the linear evolution of inhomogeneities\cite{Koivisto-2006} plus the Hubble diagram of type Ia supernovae and baryon oscillations\cite{AEMM-2006}. Besides finding that the $f(R)=R-\mu^4/R$ model was strongly disfavored by the data, it was found that the combined observational data were capable of reducing the allowed parameter space of the exponent $\beta$ to an interval of order $\sim 3 \times 10^{-5}$ around $\beta=0$, with $\alpha$ having a value similar to the cosmological constant. This meant that $R-\alpha R^\beta\approx R-\alpha-\alpha\beta\ln R$ could be restricted to a tiny region around the $\Lambda$CDM model. More stringent constraints on this model were found comparing its predictions with the CMB and matter power spectra\cite{Li:2006ag}, pushing the $\beta$ parameter to the range $\sim 10^{-6}$, thus making this model virtually indistinguishable from $\Lambda$CDM. These conclusions have been reconfirmed by considering updated data\cite{Carvalho:2008am,Santos:2008qp,Pires-2010} and strong lensing statistics\cite{YC-2009,Ruggiero:2007jr}. A different class of models\cite{Baghram}, with $f(R)=(R^n-R_0^n)^{1/n}$, has also been confronted recently with various data samples. The constraints on the parameters, $n=0.98\pm 0.08$, also place this model in the vicinity of the $\Lambda$CDM model.\\
The models considered so far modify the gravitational dynamics at late times, which turns out to be strongly constrained by observations. Modifications at early times should be very weak because of the strong constraints imposed by big bang nucleosynthesis and CMB primary anisotropies. One could thus consider whether modifications at intermediate times could be in agreement with observations. A model proposed in this direction\cite{Li-Chu-2006} takes the form $f(R)=R+\lambda_1 H_0^2 e^{-|R|/(\lambda_2 H_0)^2}$, where $H_0$ represents the current value of the Hubble parameter, $\lambda_1$ measures the magnitude of the departure from GR, and $\lambda_2$ controls the time at which the correction becomes relevant. Note that at late times this $f(R)$ recovers the $\Lambda$CDM model (which corresponds to the limits $R\to 0$ or $\lambda_2\to \infty$). Though the background evolution of this model is not significantly different from the standard $\Lambda$CDM model for $\lambda_2=500,1000$, which means that it can hardly be constrained by type Ia supernovae data, its effects on the CMB and matter power spectra are dramatic, being $\lambda_2=1000$ safely excluded. The strongest constraints are imposed by the matter power spectrum. This can be understood by looking at the growth
equation for the comoving energy density fluctuations\cite{Koivisto-2006,Koi-Kurki-2006,ULT-2007} $\delta_m$ for large momentum $k$
\begin{equation}
\frac{d^2\delta_m}{dx^2}\approx-\frac{k^2c_s^2}{a^2H^2}\delta_m \ ,
\end{equation}
where $x=\log a(t)$, and $c_s^2=\dot{f}_{R}/(3f_R(2f_R H+\dot{f}_{R}))$ represents the effective sound speed squared. If $c_s^2>0$, the perturbations oscillate instead of growing, whereas for $c_s^2<0$ they become unstable and blow up (this happens for $f(R)=R-\alpha R^\beta$ if $\beta>0$). In the 
$\Lambda$CDM model $c_s^2=0$. The form of the matter power spectrum in the exponential and power-law models, therefore, changes significantly with time developing an intricate oscillatory structure for larger $k$ that clearly conflicts with observations, which allows to strongly constrain the parameter space of the models. The most optimistic constraints restrict the parameter $\lambda_2$ to the region\cite{Li-Chu-2006} $\lambda_2\geq 5\times 10^4$. \\ 
In parallel to the considerations of above, a theoretical consistency check using phase space analysis\cite{FayTT-2006,TUT-2008} was also carried out to determine whether some families of $f(R)$ models could allow for the different phases in the expansion history of the universe suggested by observations. It was shown that radiation, matter, and de Sitter points exist irrespective of the form of the function $f(R)$ provided that the function 
\begin{equation}
C(R)=-3\frac{(Rf_R-2f)Rf_{RR}}{(Rf_{R}-f)(Rf_{RR}-f_R)}
\end{equation}
does not show discontinuous or divergent behaviors. Thus models satisfying the condition $C(R)>-3$ lead to a background evolution comprising the sequence of radiation, matter and de-Sitter epochs. From this it follows that, unlike in metric formalism, theories of the type $f(R)=R-\beta/R^n$ do allow for the sequence of radiation-dominated, matter-dominated, and de Sitter eras if $n>-1$. For theories of the type $f(R)=R+\alpha R^m-\beta/R^n$, one finds that an early inflationary epoch is not followed by a standard radiation-dominated era, which conflicts with the idea that early and late time cosmic acceleration could be unified with this type of models\cite{Sot-2005}. In particular, for $m>2$, the inflationary era is stable and prohibits the end of inflation; if $3/2<m<3$, then inflation ends with a transition to a matter-dominated phase, which is then followed by late time acceleration; for $4/3<m<3/2$, inflation is not possible; and for $0<m<4/3$ one can have the sequence of radiation-dominated, matter-dominated, and late-time de Sitter without early-time inflation. \\

\section{Solar system and laboratory tests \label{sec:local}}

Most of the $f(R)$ models found in the literature have been proposed to address phenomenological issues related with the largest scales. It is generally argued that at such scales the theory of gravity could depart from GR, implying that GR should be seen as an approximation valid only at certain scales. This, in a sense, justifies the study of $f(R)$ theories that are very far away from GR, i.e., that are not of the form $f(R)=R+$small corrections. However, this viewpoint should be supported by an explicit mechanism able to explain why/how the gravity Lagrangian should/could change its form depending on the scales involved. Since such a mechanism has not been seriously discussed in the literature on Palatini theories, we assume that the proposed $f(R)$ models should be treated in the same way as GR and, therefore, to be viable they should agree with observations and experiments on all scales. For this reason, the same $f(R)$ models that have been proposed to explain the cosmic speedup should be in accord with the dynamics of the solar system and laboratory systems. In this section we address these points in detail. 

\subsection{Solar system}

In section \ref{sec:II}, we remarked that the field equations of Palatini $f(R)$ theories in vacuum boil down exactly to those of GR with a cosmological constant. For this reason, if one considers a non-rotating, spherically symmetric star like the sun, the metric outside the star can be written as a Schwarzschild-de Sitter solution
\begin{equation}\label{eq:S-dS}
ds^2_{SdS}=g_{\mu\nu}dx^\mu dx^\nu=-A({r})d{t}^2+\frac{d{r}^2}{A({r})}+{r}^2d\Omega^2
\end{equation}
with $A({r})=1-2GM_\odot/{r}-\Lambda_{eff} {r}^2/3$, where $\Lambda_{eff}$ represents a cosmological constant, $G$ is Newton's constant, and $M_\odot$ is identified with the mass of the star. The well-known model $f(R)=R-\mu^4/R$, like any other $f(R)$ model, admits such solutions. In this particular case, it is easy to see that $\Lambda_{eff}=\sqrt{3}\mu^2/4$. One is then tempted to conclude that this model is compatible with solar system observations\cite{Vollick-2003} because for sufficiently small $\Lambda_{eff}$ its predictions are virtually indistinguishable from those of the Schwarzschild\cite{RI-Schw} and Kerr\cite{RI-Kerr} solutions of GR, which pass all observational tests. However, the situation is more subtle because, due to the modified dynamics within the sources, the transition from the interior solution to the exterior solution is not, in general, as simple as in GR. To illustrate this point, let us consider a 
presureless body such as a rocky planet or a gold sphere, for example. For such objects a formal analytical solution for an arbitrary Lagrangian $f(R)$ can be easily obtained\cite{Olmo2007b} by writing the field equations in the form (\ref{eq:Gab-fEr}) and taking the ansatz
$ds^2=g_{\mu \nu }dx^\mu dx^\nu=\phi^{-1}h_{\mu \nu }dx^\mu dx^\nu$ with
\begin{equation}\label{eq:metric-int}
ds^2=\frac{1}{\phi(T)}\left[-B(r)e^{2\Phi(r)}dt^2+\frac{1}{B(r)}dr^2+r^2d\Omega ^2\right] \ ,
\end{equation}
where we have defined $\phi(T)\equiv \frac{f_R(0)}{f_R(T)}$ to guarantee that outside the sources $g_{\mu\nu}=h_{\mu\nu}$ (see the discussion above Eq. (\ref{eq:Gab-fEr})). We then find 
\begin{eqnarray}\label{eq:Phi0}
\frac{2}{r}\frac{d\Phi}{dr}&=&\frac{\tilde{\kappa}^2}{\phi^2}\left(\frac{{T}_r^r-{T}_t^t}{B}\right) \\
-\frac{1}{r^2}\frac{d(r[1-B])}{dr}&=& \frac{\tilde{\kappa}^2{T}_t^t}{\phi^2}-\tilde{\Lambda}(T)\label{eq:B}
\end{eqnarray}
Defining now $B(r)=1-2\tilde{G} M(r)/r$ in (\ref{eq:B}),
we can rewrite $M(r)$ and $\Phi(r)$ as
\begin{eqnarray}\label{eq:M}
M(r)&=&-\frac{\tilde{\kappa}^2}{2\tilde{G}}\int_0^r dx \ x^2 \left[\frac{T_t^t}{\phi^2}-\tilde{\Lambda}(T)/\tilde{\kappa}^2\right]\\
\Phi(r)&=&\frac{\tilde{\kappa}^2
}{2}\int^r_0dx \ x \left[\frac{{T}_r^r-{T}_t^t}{\phi^2B}\right]
\label{eq:Phi}
\end{eqnarray}
If we consider a point outside of the sources at radius $r$, the above equations can be readily integrated leading to
\begin{eqnarray}\label{eq:M-ext}
M(r)&=&M_\odot+\frac{\tilde{\Lambda}(0)}{6\tilde{G}}r^3\\
\Phi(r)&=&\Phi_0 \label{eq:Phi-ext} \ , 
\end{eqnarray}
where $\tilde{\Lambda}(0)=f(0)/2f_R(0)$, $M_\odot$, and $\Phi_0$ are constants. Since we are assuming a presureless fluid, $T_t^t=-\rho$, taking units such that $\tilde{\kappa}^2=8\pi \tilde{G}$ we find that $M_{\odot}$ and $\Phi_0$ are given by
\begin{eqnarray}\label{eq:Modot}
M_\odot &=&\int_0^{R_\odot} dr \frac{4\pi r^2\rho}{\phi^2(T)}+\frac{1}{2\tilde{G}}\int_0^{R_\odot} r^2dr \tilde{\Lambda}(-\rho) \\
\Phi_0 &=& \tilde{G}\int_0^{R_\odot} dr \frac{4\pi r^2\rho}{\phi^2(r-2\tilde{G}M(r))}
\end{eqnarray}
where $R_\odot$ is the radius of the object (where $\rho$ vanishes). 
With these results, outside of the sources (where $\phi(0)=1$) the line element (\ref{eq:metric-int}) coincides with (\ref{eq:S-dS}) if we absorb the constant factor $e^{2\Phi_0}$ in a redefinition of the time coordinate, identify $A(r)$ with $B(r)$, and take $\Lambda_{eff}= \tilde{\Lambda}(0)$.\\
Once an $f(R)$ Lagrangian  is specified, eqs. (\ref{eq:metric-int}), (\ref{eq:M}), and (\ref{eq:Phi}) provide a complete exact solution for a presureless, nonrotating, spherical object. The usual GR expressions are recovered by just taking $\phi=1$ and $\tilde{\Lambda}=0$. In particular, one finds the Newtonian expression for the mass $M_\odot=\int_0^{R_\odot} dr 4\pi r^2\rho$. We can thus use this general solution to study the Newtonian limit corresponding to such objects. We start by writing down the general expression for the $g_{tt}$ component of the metric 
\begin{equation}\label{eq:gtt}
g_{tt}=-\frac{1}{\phi(T)}\left(1-\frac{2\tilde{G}M(r)}{r}\right)e^{2(\Phi(r)-\Phi_0)}
\end{equation}
 If we consider, for instance, the $f(R)=R-\mu^4/R$ theory, using (\ref{eq:R-CDTT}) we can write
\begin{equation}\label{eq:phi-1/R}
\phi(T)=1-\frac{1}{2\left[1+\sqrt{1+12\left(\frac{\rho_\mu}{\rho}\right)^2}\right]} \ .
\end{equation}
\begin{figure}[ht]
\begin{center}
\psfig{file=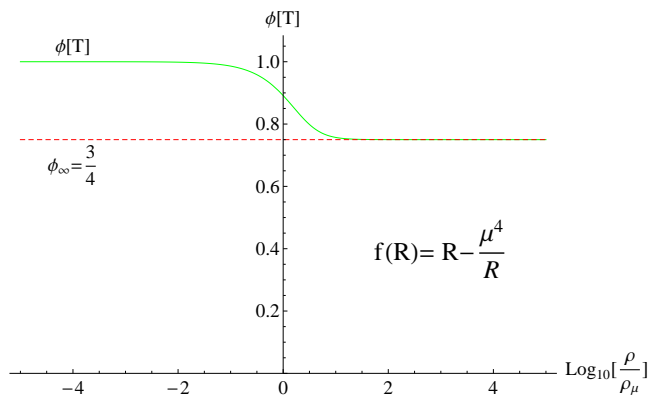,width=8.5cm}
\vspace*{8pt}
\caption{Dependence of $\phi(T)\equiv f_R(T)/f_R(0)$ on $\rho/\rho_\mu$ in the $1/R$ model. The function smoothly interpolates between the two asymptotic constants $\phi_\infty=3/4$ and $\phi_0=1$.\label{phi}}
\end{center}
\end{figure}
From this expression we see that $\phi(T)$ varies continuously from $\phi_{\infty}=3/4$ inside matter ($\rho\gg \rho_\mu$) to $\phi_0=1$ in vacuum. This should have disastrous consequences for the theory because in the solar system the $g_{tt}$ component of the metric is only slightly different from unity, with the largest corrections being of order $\tilde{G}M_\odot/R_\odot\sim 10^{-5}$ near the Sun. The amplitude of the change in $\phi(T)$ for the $1/R$ model implies a change in the metric of order $\sim 1/3$, which is comparable to the change in the metric when going from infinity to nearby the event horizon of a black hole. The difference is that this variation in the metric occurs in a much shorter scale, which must produce even larger accelerations.  Something similar happens to the models $f(R)=R-\mu^{2(n+1)}/R^n$, for which the change in $\phi$ is of order $\sim n/(n+2)$. To save those models, one could argue\cite{Sot-2006,TUT-2008} that the density in the solar system is always much larger than $\rho_\mu$, which could prevent $\phi(T)$ from reaching its vacuum value $\phi_0$. However, this seems a very weak argument because the structure of matter is discrete (localized wavepackets) and, therefore, one can always find regions in which $\phi(T)$ takes all possible values, which should have observable consequences at microscopic scales. We will see later a detailed example of this in Sec. \ref{sec:atoms}. For the moment, we just conclude that to have a chance of being viable according to local experiments, any Palatini $f(R)$ theory must be characterized by a function $f(R)$ such that $f_R(T)$ is not very sensitive to density variations over the range of densities accessible to those experiments. From a technical point of view, this simply means that $\phi(T)$ must be almost constant because then with a simple constant rescaling of the metric one can bring $\phi(T)$ from $\phi(T)\approx \phi_0+$corrections to $\tilde{\phi}(T)\approx 1+$corrections, which turns the metric into its standard almost-Minkowskian form $g_{\mu\nu}=\eta_{\mu\nu}+$corrections. In particular, if one accepts that local experiments are carried out in an environment with density $\rho_{local}\gg \rho_\mu$, for the $1/R$ model $\tilde{\phi}$ would be\footnote{This new constant rescaling of the metric is equivalent to using the same system of units in Einstein and Jordan frames in regions where $\rho\gg\rho_\mu$.} $\tilde{\phi}\equiv f_R(T)/f_R(\infty)= f_R(T)=4\phi(T)/3$. \\

Since for viable models we must have $\tilde{\phi}(T)\approx 1+\Omega(T)$, with $|\Omega(T)|\ll 1$, we can express the metric outside  spherical bodies as 
\begin{equation}\label{eq:g_tt}
g_{tt}\approx -1+2U+\Omega(T) \ , 
\end{equation}
where $U=\tilde{G}M_\odot/r+\Lambda_{eff}r^2/6$, and $\Omega(T)$ is sensitive to the sources present at radius $r$. This unusual local dependence must also be very weak, which can be used to impose constraints on the family of allowed Lagrangians. A detailed discussion of such constraints can be found in Ref.~\refcite{Olmo2005}, where the Newtonian and post-Newtonian limits of $f(R)$ theories in metric and Palatini formalism was worked out. The results obtained in Ref.~\refcite{Olmo2005} using perturbative methods coincide with the Newtonian expansion from the exact solutions given here\footnote{In Ref.~\refcite{Olmo2005} there seems to be a wrong sign in front of $\Lambda_{eff}$. That error seems to be a transcription error because the perturbation equations of the Appendix have the right sign (compare them with the metric case). In any case, that sign is irrelevant for the conclusions of that work.}, which provides an independent confirmation of their validity (up to Newtonian order at least) without the complexities that the post-Newtonian analysis involves. Other approaches to the Newtonian limit of Palatini $f(R)$ theories\cite{Meng-Wang-Newton,DomBarraco,AFRT-2005} have reported the existence of Yukawa-type corrections to the usual Newtonian potential with a length scale of order $l\sim 1/\sqrt{\Lambda_{eff}}$. However, such terms are generically associated with propagating scalar degrees of freedom, which do not exist in the Palatini version of $f(R)$ theories. Moreover, those Yukawa-type corrections should also appear in the post-Newtonian parameter $\gamma$, as it happens in the metric version of $f(R)$ theories and scalar-tensor theories, which would be in conflict with experiments due to the long interaction range $l\sim 1/\sqrt{\Lambda_{eff}}$. Additionally, in the non-perturbative exact solution derived here and in the perturbative approach of Ref.~\refcite{Olmo2005} there is no trace of such Yukawa-type correction. A correcting term of the form $A\rho$ similar to the one denoted here\footnote{Note that if $\tilde{\phi}(T)$ admits a perturbative expansion, then $\tilde{\phi}(T)\approx 1+T\partial_T\tilde{\phi}$, which in the Newtonian limit implies that $\Omega(T)=-\rho\partial_T\tilde{\phi}$. } by $\Omega(T)$ accompanying the Newtonian potential in (\ref{eq:g_tt}) was also found in Refs.~\refcite{DomBarraco,AFRT-2005} but not in Ref.~\refcite{Meng-Wang-Newton}. An interesting interpretation of that term in the particular model $f(R)=R+\lambda^2R^2$ can be found in Ref.~\refcite{BGHV-1996}, where that Palatini model was compared with its metric version. The metric version has a Yukawa-type correction of the form $\Delta V\propto\lambda^{-2}\int d^3\vec{x} \rho(\vec{x})e^{-|\vec{x}-\vec{x}_0|/\lambda}/|\vec{x}-\vec{x}_0|$, which for very short interaction range, $\lambda\to 0$, leads to  $\Delta V\propto \rho(\vec{x}_0)$. This allows to see the density-dependent term as the limiting case of a Yukawa interaction when the interaction range is ultra-short. A similar interpretation is possible for the behavior of the scalar curvature in the metric version of general $f(R)$ theories\cite{Olmo2007a}. 
 We also mention that similar expressions for the non-perturbative eqs. (\ref{eq:metric-int}), (\ref{eq:M}), and (\ref{eq:Phi}) have also been found in Refs.\refcite{BH-2000,BB-2007}. Though from the definitions in those works, $B(r)\equiv e^{-\alpha(r)}$ and $B(r)e^{2\Phi(r)}\equiv e^{\gamma(r)}$, one finds exact agreement with our results, an erroneous identification in eq. $(36)$ of Ref.\refcite{BH-2000} (or equivalently eq. $(19)$ of Ref.\refcite{BB-2007}) leads to different conclusions. To be precise, they claimed that $e^\gamma=e^{-\alpha}+e^{2\Phi}$, whereas from their field equations (and ours) one finds $e^\gamma=e^{-\alpha+2\Phi}$. The discussion of the Newtonian limit given in Ref.\refcite{BB-2007} also assumes that $f(R(T))$ can be expanded around $T=0$ as $f(T)\approx f(0)-\rho\partial_T f|_0\ldots $, which for models such as the $1/R$ (see eq. (\ref{eq:phi-1/R})) is not justified. \\
 
Following Ref.\refcite{Olmo2005} and our previous discussion, one finds that the weak dependence of $\phi(T)$ on the density implies that a change $\Delta{\phi}$ relative to the value of ${\phi}$ induced by a change $\Delta\rho$ relative to the density $\rho$ must be very small, which can be expressed as 
\begin{equation}\label{eq:Newtonian}
\left|\frac{\rho}{f_R}\frac{\partial f_R}{\partial \rho}\right|\ll 1 \ \  \leftrightarrow \ \ \left|\frac{{\kappa}^2\rho}{R f_R}\right|\left|\frac{1}{1-f_R/Rf_{RR}}\right|\ll 1 \ ,
\end{equation}
This result together with the fact that $R\approx \kappa^2\rho$ and $f_R(T)\approx 1 $ in local experiments, can be reduced\cite{TUT-2008} to the condition
\begin{equation}
|R f_{RR}|\ll1 \ .
\end{equation}
When applied to models of the form $f(R)=R-\mu^{2(n+1)}/R^n$, one finds that 
\begin{equation}
|Rf_{RR}|=\left|n(n+1)\left(\frac{\mu^2}{R}\right)^{n+1}\right|\approx \left|n(n+1)\left(\frac{\rho_\mu}{\rho_{local}}\right)^{n+1}\right| \ ,
\end{equation}
which is much smaller than unity if $n>-1$ as long as $\rho_{local}\gg\rho_\mu\sim 10^{-26}$ g/cm$^3$. This argument and others similar to this have been used in the literature to claim that this family of models are not very constrained by local experiments, which justified their cosmological analysis\cite{TUT-2008,Sot-2006}. However, as we pointed out above, it relies on the assumption that the density scale $\rho_\mu$ is not reachable under regular experimental conditions. By considering microscopic experiments, we will show next that this assumption is not correct. Local experiments, therefore, will be able to test the Palatini $f(R)$ dynamics and impose tight constraints on the family of allowed models. \\

\subsection{Microscopic experiments}\label{sec:atoms}

Shortly after Vollick's proposal for explaining the cosmic speedup using the Palatini version of the $1/R$ model, it was claimed\cite{Flanagan-2004a} that the model was in conflict with electron-electron scattering experiments. The argument goes as follows. Since the affine connection can be expressed in terms of the metric and the matter sources according to (\ref{eq:LC-h}), by inserting back this solution into the action, one ends up with a theory that has new interactions among matter fields and between the matter fields and the curvature. The original discussion of this problem was carried out in the Einstein frame representation of the theory, which apparently allows for a simpler interpretation of the action
\begin{equation} \label{eq:ST-actionE}
S=\frac{1}{2\kappa^2}\int d^4
x\sqrt{-{h}}\left(R({h})-\frac{V(\phi)}{\phi^2(T)}
\right)+S_m[\phi^{-1}(T)h_{\mu \nu},\psi_m] \ ,
\end{equation}
where $\phi=\phi(T)$ is, in general, given by solving (\ref{eq:phi-ST-Pal}), and in our particular case takes the form (\ref{eq:phi-1/R}) with $\rho$ replaced by $T$. The explicit coupling to the matter of the factor $\phi(T)$ in $S_m$ together with the new matter term $\frac{V(\phi)}{\phi^2(T)}$ were inmediately interpreted as a clear indication that the theory should be in conflict with particle physics experiments. This view was seriously criticized\cite{Vollick-2004} (see also Ref. \refcite{Vollick-2005}) because if the theory is analyzed in the original Jordan frame, the direct coupling of $\phi(T)$ with the matter action disappears and, on grounds of the Einstein equivalence principle, no new effects should be observed in a freely falling frame. This observation raised (again) a debate on the mathematical and physical consequences of working with different field redefinitions and/or frames. A reanalysis of the problem\cite{Flanagan-2004b} then put forward the existence of non-perturbative couplings that prevented a consistent perturbative treatment of some of the new interaction terms. The theory thus seemed intractable and a definitive conclusion about its viability could not be explicitly worked out.\\

The problem of the influence of the $f(R)$ dynamics in microscopic systems was reconsidered\cite{Olmo-2008a} with the focus on the possible effects  that the gravitationallly-induced matter interactions could have on the stability of Hydrogen. Starting with the equation for a Dirac field in curved space\footnote{This analysis was carried out in the original Jordan frame to avoid the discussion generated by the use of the Einstein frame variables.}, the corresponding non-relativistic Schr\"{o}dinger equation for an electron in an external electromagnetic field was derived. The goal was to study if the density-dependent function $\phi(T)$ present in the metric (\ref{eq:metric-int}) could have an effect on atoms.
This question is pertinent because atoms are systems in which the matter density is localized around the nucleus and drops to zero as we move away from it. Since the density scale $\rho_\mu$ will be reached at some point, the stability and structure of atoms provides a natural laboratory to test the Palatini dynamics of theories sensitive to very low density scales.  The fact that the average distance between atoms in a diluted perfect fluid is much larger than their typical sizes guarantees that they can be seen as isolated systems immersed in a perfect vacuum, which provides a way out of the problem posed by the belief that local experiments are carried out within an environment whose density hides the presence of the modified dynamics. \\
 Neglecting the Newtonian potential corrections, the metric (\ref{eq:metric-int}) boils down to $g_{\mu\nu}=\phi^{-1}\eta_{\mu\nu}$. For a metric of this type, one finds the following Schr\"{o}dinger-Pauli equation
\begin{eqnarray}\label{eq:Pauli}
E\eta&=& \left\{\frac{1}{\tilde{m}+m_0}[(\vec{p}-e\vec{A})^2-e\vec{\sigma}\cdot\vec{B}]+eA_0\right\}\eta \\&+&\left\{\frac{1}{\tilde{m}+m_0}\left[i\vec{\sigma}(\vec{\nabla}\Omega\times\vec{\nabla})-2ie(\vec{A}\cdot\vec{\nabla}\Omega)+\vec{\nabla}^2\Omega-|\vec{\nabla}\Omega|^2+2(\vec{\nabla}\Omega \cdot\vec{\nabla})\right]+(\tilde{m}-m_0)\right\}\eta  \nonumber
\end{eqnarray}
where $E$ is the non-relativistic energy of the electron, $\eta$ is a two-component spinor, $\tilde{m}\equiv m \phi^{-1/2}$, $m_0$ is a constant of order the mass of the electron $m$, $\vec{\sigma}$ are the Pauli matrices, $A_0$ and $\vec{A}$ are the components of the electromagnetic potential $4$-vector, $\vec{B}$ is the external magnetic field,  $\Omega=\frac{3}{4}\ln \phi(T)$, and $T=-m\eta^\dagger \eta$. If one considers GR, $\phi=1$, the usual Schr\"{o}dinger-Pauli equation is recovered by just identifying $m_0$ with $m$. For concreteness, let us consider the $1/R$ model\cite{Vollick-2003}. In this case, the function $\phi(T)$ is given in (\ref{eq:phi-1/R}), and expressing length units in terms 
of the Bohr radius $a_0=0.53\times 10^{-10}$ m, we find that $\rho/\rho_\mu=10^{24}P_e(x)$, where $P_e(x)=\eta^\dagger\eta$ is the probability density of finding an electron. This expression for  $\rho/\rho_\mu$ indicates that the electron reaches the characteristic cosmic density, $\rho/\rho_\mu\approx 1$, in regions where the probability density is near $P_e(x)=10^{-24}$. In ordinary applications, one would say that the chance of finding an electron in such regions is negligible. In our case, however, that scale defines the transition between the high density $\rho\gg \rho_\mu$ and the low density $\rho\ll \rho_\mu$ regions. In regions of high density, one finds that $\phi$ rapidly tends to a constant, $\phi_\infty=3/4$, which leads to $\tilde{m}=2m/\sqrt{3}$ and $\vec{\nabla}\Omega=0$. Identifying $m\to \sqrt{3}m_0/2$, Eq. (\ref{eq:Pauli}) reduces to the usual Schrodinger-Pauli equation
\begin{equation}\label{eq:Sch-Pauli}
\E\eta= \left\{\frac{1}{2m_0}[(\vec{p}-e\vec{A})^2-e\vec{\sigma}\cdot\vec{B}]+eA_0\right\}\eta 
\end{equation}
In regions of low density, $\phi$ tends to unity, $\vec{\nabla}\Omega=0$, and $\tilde{m}\to m$ as $\rho/\rho_\mu\to 0$. As a result, the mass factor dividing the kinetic term is now a bit smaller ($m_0>m$) than in the high density region, but the mass difference $\tilde{m}-m_0$ is no longer zero. This is the crucial point, because $\tilde{m}-m_0\approx -0.13m_0$ represents a deep potential well in the outermost parts of the atom (from $r\approx 26a_0$ to infinity), which has important consequences for its stability. In fact, if one assumes that the electron is initially in the ground state, the deep potential well that appears in the outer regions of the atom makes this state unstable and triggers a flux of probability density (via quantum tunneling) to those regions. Using time dependent perturbation theory, the half life of Hydrogen subject to this potential turns out to be
\begin{equation}\label{eq:halflife}
\tau_H\equiv\frac{\hbar}{\Gamma}\approx 6\cdot 10^3 s \ ,
\end{equation}
which is in clear conflict with observations. From this analysis we extract several lessons. First, we have shown that the ultra low density scales that characterize models aimed at explaining the cosmic speedup can be reached in microscopic scenarios. Second, we have seen that the modified gravitational dynamics of those models can have nontrivial effects on systems such as the Hydrogen atom. Third, we can use standard perturbative techniques to estimate those effects and constrain the models. The results obtained here for the $1/R$ model also provide a simple test to determine whether a given model is compatible with observations or not. Since the instability of the ground state is to a large extent due to the potential well induced by the mismatch between the values of $\tilde{m}$ and $m_0$ in the low density regions, any $f(R)$ model that yields a non-negligible difference 
\begin{equation}\label{eq:Deltam}
\Delta_m=m_0\left(\sqrt{\frac{f_R(\infty)}{f_R(0)}}-1\right)
\end{equation}
can be automatically ruled out. In particular, for the family $f(R)=R-\mu^{2(n+1)}/{R^n}$ we find that $f_R(\infty)=1$ and $f_R(0)=1+n/(n+2)$, which leads to $\Delta_m=m_0\left(\sqrt{\frac{1}{1+\frac{n}{n+2}}}-1\right)$. This quantity is small only if $|n|\ll 1$, which yields $\Delta_m\approx -m_0 n/4$. For not too small $n>0$, the results of Ref.\cite{Olmo-2008a} could be directly used to estimate the half-life of the atom. However, for very small $n$, the estimation of the half-life should be reconsidered taking into account the contributions coming from the $\nabla\Omega$ terms, which were negligible transient potentials in the $1/R$ case. For negative values of $n$ one should note that rather than a potential well, one finds a potential barrier, which would lower the energy of the ground state. In any case, all these possibilities have observable effects and could be strongly constrained with current data. In this sense, the variation in the energy levels of Hydrogen has been estimated\cite{LMS-2008,LMS-2009} for models in which the condition $\Delta_m$ is satisfied, i.e., for models of the form $f(R)=R(1+\epsilon(R))$ such that $|\epsilon(R)|\ll 1$. Since that analysis was carried out in the Einstein frame and with a non-linear redefinition of the Dirac field, we will refrain ourselves from giving a detailed correspondence between our formulas and those. The strategy followed there to constrain the models consisted on determining how the energy of a photon released due to one transition changes relative to that emitted in another transition, which yields a quantity that is independent of the electron mass and also seems to be independent of the choice of frame. Using data for the transitions from the initial state $(n,l)=(2,0)$ to the final state $(n,l)=(1,0)$, and from $(n,l)=(8,3)$ to $(n,l)=(2,0)$, the following constraint was found
\begin{equation}
\left|\frac{f_{RR}H_0^2}{f_R}\right|\leq 4\times 10^{-40}
\end{equation}
For the family of models $f(R)=R-\mu^{2(n+1)}/{R^n}$, this constraint implies\footnote{To obtain this result we consider as valid the assumptions made in Ref.\refcite{LMS-2008,LMS-2009}, evaluate $f_R$ and $f_{RR}$ in the vacuum value $R_{vac}=(n+2)^{1/(n+1)}\mu^2$, and approximate $H_0^2$ by $H_0^2=\mu^2(n+2)^{1/(n+1)}/12$.} that $|n|\leq 10^{-38}$. This is the tightest constraint put so far on this family of models (recall that from CMB anisotropies and baryon oscillations the bound was around $|n|\leq 10^{-6}$) and puts forward the relevance of microscopic experiments for the understanding of the dynamics of Palatini theories. \\

The analysis of Ref.\refcite{Flanagan-2004a} also raised doubts about the applicability of the Palatini field equations to describe macroscopic systems. A careful analysis of such problem has been explicitly carried out in Refs.\refcite{LMS-2008,LMS-2009} (see also Ref.\refcite{SAS:2009} for a related discussion). It was concluded there that at the classical level the physical masses and geodesics of particles, cosmology, and astrophysics in Palatini modified gravity theories are all indistinguishable from the results of general relativity plus a cosmological constant. Part of this argument was supported by the assumption that isolated particles are stable and should not exhibit violations of energy and momentum conservation. Though this could be true for certain Palatini models, the stability of microscopic systems can not be guaranteed in general. In particular, it is in clear conflict with the results presented here for the Hydrogen atom and the family of models $f(R)=R-\mu^{2(n+1)}/R^n$. Since there is a flux of probability density to infinity, the energy and momentum of the system are not locally conserved. Thus, though GR is holographic in the sense that the equations of motions for a localized distribution of energy and momentum surrounded by vacuum can be derived by considering surface, rather than volume, integrals over curvature components, the instability of certain isolated systems in some Palatini $f(R)$ models may prevent the interpretation of the {\it exterior} space-time as completely vacuum and, therefore, as not exactly equivalent to that of GR plus a cosmological constant. To the light of this, we believe that part of the conclusions of Refs.\refcite{LMS-2008,LMS-2009} should be reconsidered.  

\section{Other tests \label{sec:others}}

The previous sections have provided us with a good deal of information about the properties of Palatini $f(R)$ theories. We have contrasted the dynamics of these theories against cosmological, solar system, and laboratory data, and this has allowed us to impose severe constraints on the form of some families of $f(R)$ Lagrangians. This exercise has been particularly useful for constraining models characterized by ultralow density scales. We now review other approaches followed in the literature to understand the viability and robustness of $f(R)$ theories and which have raised interesting debates. We will begin by considering the structure of spherically symmetric, static stars and then will focus on the initial value formulation of these theories. 

\subsection{Stellar structure}

In this section we consider a problem that initially seemed to affect seriously the theoretical viability of {\it all} $f(R)$ models in Palatini formalism. Using the Tolman-Oppenheimer-Volkov (TOV) equations for the interior of stars in equilibrium\cite{Kainu07a}, it was found\cite{Sot08a} that certain terms in those equations could blow up and form curvature singularities near the surface of spherically symmetric, static matter configurations with polytropic equation of state, $\rho(P)=(P/K)^{1/\gamma}$, with index $3/2<\gamma<2$. Since the physically interesting case $\gamma=5/3$ (degenerate, non-relativistic fermion gas) lies within this interval, this result was regarded as a serious theoretical concern about the viability of Palatini $f(R)$ theories. The problem was soon reconsidered\cite{Kainu07b} and interpreted differently, claiming that it had more to do with the  peculiarities of the equation of state used than with the own structure of Palatini $f(R)$ theories. This was based on the observation that for neutron stars the tidal acceleration due to the surface singularity becomes equal to the Schwarzschild value of GR only at a distance  $\sim 0.3$ fermi from the surface of the star, which makes unrealistic the use of a polytropic equation of state. However, this conclusion was also challenged\cite{Sot08b} claiming that the fluid approximation is still valid on the scales at which the tidal forces diverge just below the surface of a polytropic sphere in the case of the generic functions $f(R)$ considered. This debate was independently reexamined \cite{Olmo-2008b} reaching an intermediate answer, which is the one that we present here. \\

Consider a static, spherical object described by a perfect fluid, with $T_{\mu\nu}=(\rho+P)u_\mu u_\nu+P g_{\mu\nu}$. Take Schwarzschild-like coordinates, and parametrize the space-time line element as
\begin{equation}
ds^2=-A(r)e^{2\psi(r)}dt^2+\frac{1}{A(r)}dr^2+r^2d\Omega^2 \ ,
\end{equation}
where $A(r)=1-2M(r)/r$. Inserting these inputs in the field equations (\ref{eq:Gab-f}), one obtains the following TOV coupled equations\footnote{These equations correct some transcription errors present in Ref.\refcite{Olmo-2008b}.}
\begin{eqnarray}\label{eq:psir-fin}
\left(\frac{f_{R,r}}{f_R}+\frac{2}{r}\right)\psi_r &=&\frac{\kappa^2(\rho+P)}{f_RA}
-\frac{3}{2}\left(\frac{f_{R,r}}{f_R}\right)^2+\frac{f_{R,rr}}{f_R}
\\
\label{eq:Mr-fin}
\left(\frac{f_{R,r}}{f_R}+\frac{2}{r}\right)\frac{M_r}{r}&=&\frac{f+\kappa^2(\rho+3P)}{2f_R}+A\left[\frac{f_{R,rr}}{f_R}+\frac{f_{R,r}}{f_R}\left(\frac{2r-3M}{r(r-2M)}-\frac{3}{4}\frac{f_{R,r}}{f_R}\right)\right]\\
P_r&=&-\frac{P^{(0)}_r}{[1-\alpha(r)]}\frac{2}{\left[1+\sqrt{1-\beta(r)P^{(0)}_r}\right]} \label{eq:pressure-fin}
\end{eqnarray}
where $f_{R,r}\equiv \partial_r f_R$, and we have defined 
\begin{eqnarray}
P^{(0)}_r&=& \frac{(\rho+P)}{r(r-2M)}\left[M-\left(\frac{f+\kappa^2(P-\rho)}{f_R}\right)\frac{r^3}{4}\right] \\
\alpha(r) &=& (\rho+P)\frac{f_{R,P}}{f_R}\\
\beta(r) &=& (2r)\frac{f_{R,P}}{f_{R}}\left[1-\frac{3(\rho+P)}{4}\frac{f_{R,P}}{f_{R}}\right]
\end{eqnarray}
with $f_{R,P}\equiv \partial_P f_R$. One can check that these expressions recover the GR formulas in the limit $f_R=1$ and $f=R-2\Lambda$. Given an equation of state, $P=P(\rho)$ or $\rho=\rho(P)$, one can use the above formulas to compute the structure of static, spherically symmetric objects. To do it, one must first express the functions $f$ and $f_R$ in terms of $T=-\rho+3P$ and rewrite the radial derivatives of $f_R$ in the form $f_{R,r}=f_{R,P}P_r$ and $f_{R,rr}=f_{R,P} P_{rr}+f_{R,PP}P_r^2$. One then finds
\begin{eqnarray}\label{eq:f_P}
f_{R,P}&\equiv& \frac{\kappa^2 f_{RR}}{R f_{RR}-f_R}\left(3-\rho_P\right)\\
f_{R,PP}&=&-\frac{\kappa^4 f_R f_{RRR}}{(R f_{RR}-f_R)^3}\left(3-\rho_P\right)^2-\frac{\kappa^2 f_{RR}}{(R f_{RR}-f_R)}\rho_{PP} \label{eq:f_PP} \ ,
\end{eqnarray}
where $\rho_P\equiv \frac{d\rho}{dP}$ and $\rho_{PP}\equiv \frac{d^2\rho}{d^2P}$. The terms $\rho_P$ and $\rho_{PP}$ are the reason for the existence of divergences near the surface of polytropes with index $3/2<\gamma<2$. This can be easily seen as follows. Since polytropes are characterized by $\rho(P)=(P/K)^{1/\gamma}$, one finds that $\rho_P=\rho/(\gamma P)$, and $\rho_{PP}=(1-\gamma)\rho/(\gamma P)^2$, which implies that $\rho_P$ and $\rho_{PP}$ diverge as $P\to 0$ if $\gamma>1$ in the first case and if $\gamma>1/2$ in the second case. Therefore, if those terms do not appear in the equations multiplied by appropriate powers of the pressure, divergences will be unavoidable for some values of $\gamma$. Let us now determine the dependence on $P$ of the various terms involved in the equations. From their definitions, it is easy to see that $P^{(0)}_r\sim P^{1/\gamma}$, $\alpha(r)\sim P^{-1+2/\gamma}$, and $\beta(r)P^{(0)}_r\sim P^{2(-1+2/\gamma)}$. If $\gamma<2$, those terms decay as $P\to 0$ yielding $P_r\sim P^{1/\gamma}$, but if $\gamma>2$ then they grow. The combined result for  $\gamma>2$ gives $P_r\sim P^{2-3/\gamma}$, which also falls to zero near the surface. By direct computation one also finds that $P_{rr}$, $f_{R,P}P_{rr}$, and $f_{R,P}P_{r}$ are well behaved as $P\to 0$ for $\gamma<2$. However, the term $f_{R,PP}P_r^2$ contained in $f_{R,rr}$ generates a term of the form $\rho_{PP}P_r^2\sim P^{-2+3/\gamma}$, which diverges as $P\to 0$ for $\gamma>3/2$ and produces the singularities reported in Refs.\refcite{Sot08a,Sot08b}.\\  
Contrary to the opinions provided in Refs.\refcite{Sot08a,Sot08b,Bar08}, we believe that the divergences that we have found here are not due to the differential structure of Palatini $f(R)$ theories. The fact that, unlike in GR, the field equations contain derivatives of the matter fields (via the trace $T$) up to second order is not the reason for the existence of these divergences. To see this, one should note that, 
as pointed out in Ref.\refcite{Kainu07b}, the divergent behavior of the term $\rho_{PP}P_r^2$ could be cured by simply smoothing the behavior of $\rho_P$ and $\rho_{PP}$ in the outer regions of the star using a different equation of state. Should the divergences exist even for regular equations of state, then one could blame the Palatini $f(R)$ framework for this problem but, in our case, the field equations are simply collaborating with the polytropic equation of state to the development of those infinities and, therefore, they are not directly responsible for the existence of those divergences. One should have in mind that the equations of state usually respond to statistical descriptions and involve a number of simplifying assumptions. In fact, an accurate equation of state at laboratory densities is very complicated to derive, because electrostatic interactions and other subtle effects mask the simpler statistical properties of the idealized Fermi gas approximation\cite{ST-1983}. The polytropic equation of state should therefore not be used beyond its expected regime of validity. This regime, however, may depend on the parameters that characterize the particular Lagrangian $f(R)$ considered. For instance, if one takes the model $f(R)=R\pm \lambda R^2$ with $\lambda$ of order the Planck length squared $\lambda\sim l_P^2$, which defines a density scale $\rho_\lambda\equiv (\kappa^2\lambda)^{-1}\sim 2\cdot10^{92}$ $g/cm^3$, for a neutron star the divergent term begins to be non-negligible at a density of order\cite{Olmo-2008b} $\rho_s= \left(\frac{K^2\rho_\lambda}{c^4}\right)^{\frac{1}{3-2\gamma}}\sim 10^{-210} \ g/cm^3$, which is well below any physical density one can imagine\footnote{For a free electron whose wave function is spread over the entire universe, the ratio $m_{e}/R^3_{Univ}$ is of order $\sim 10^{-118} \ g/cm^3$. Therefore, a simple electron would be enough to remove all the singularities of this type in the universe.}. But if one uses a length scale much larger than $l_P$, the terms responsible for the divergences could begin to grow in regions where the polytropic equation of state may still be valid\cite{Sot08b,Bar08}. In this sense, polytropes could still be used as a theoretical laboratory to constrain the parameters of $f(R)$ models\cite{Reij-2009}. 

\subsection{The Cauchy problem}

A very natural requirement of any theory of classical physics is that a sufficient set of initial data should be enough to determine the subsequent evolution. One then says that a theory possesses an {\it initial value formulation} if appropriate initial data (perhaps subject to constraints) can be specified such that the dynamical evolution is uniquely determined. If small changes in the initial data induce small changes in the solution over a compact region of spacetime and if such changes do not produce any changes in the solution outside the causal future of this region, then the initial value formulation is said to be {\it well posed}\cite{Wald1984}. GR has a well-posed initial value problem, which results in a stable theory with a robust causal structure. Do Palatini $f(R)$ theories have a well-posed initial value formulation? Recent works\cite{Far07,FarSot08} have concluded that, unlike their metric version, Palatini $f(R)$ theories are in general neither well-formulated nor well-posed, which seems a very serious reason for concern. We will see next, however, that Palatini $f(R)$ theories do admit in general a well-formulated initial value problem. We will also use those results to argue that the initial value problem is likely to be well-posed.\\ 

\subsubsection{Hamiltonian formulation}
  
To show that the initial value problem of Palatini $f(R)$ theories is well formulated, we consider the Brans-Dicke representation of Palatini theories and work out its $3+1$ Hamiltonian description\cite{OSA-2011} in the usual way\cite{MTW,Witten1962} (from now on we use lower-case latin letters to represent space-time indices). Consider a foliation of the spacetime manifold $\mathcal{M}$ into hypersurfaces $\Sigma_T$ of simultaneity characterized by a function $T(x)=$constant, a normalized timelike co-vector $n_a\propto \partial_a T$ normal to this hypersurface, and a {\it shift} vector $N^a$ orthogonal to $n^a=g^{ab}n_b$. This allows us to construct a time flow vector $t^a=N n^a+N^a$, where $N$ is known as {\it lapse}, and decompose the metric in the form $g_{ab}=h_{ab}-n_a n_b$. 
Elementary, though lengthy, manipulations allow us to express the Lagrangian density of (\ref{eq:ST}) as follows
\begin{eqnarray}\label{eq:Lagrangian}
\mathcal{L}&=&\frac{\sqrt{h}}{2\kappa^2}\left\{N\phi\lp R^{\lp3\rp}+(K_{ab}K^{ab}-K^2)\rp+2h^{ab}D_aND_b\phi\right.\nonumber\\
 & &\left.-\frac{\omega}{N\phi}\lp N^2h^{ab}D_a\phi D_b\phi-\lp\dot{\phi}-N^aD_a\phi\rp^2\rp\right.\nonumber\\
 & &\left.-2K\lp\dot{\phi}-N^aD_a\phi\rp-NV\lp\phi\rp\right\}
\end{eqnarray}
where $K_{ab}=h_a^c h_b^d\nabla_d n_c$ is the extrinsic curvature, $D_a\phi=h_a^b\nabla_b\phi$, $\dot\phi=t^a\partial_a\phi$, $^{(3)}R$ is the Ricci scalar of the $3-$metric $h_{ab}$, and we have used the following relations
\begin{eqnarray}
R^{(4)}&=&R^{(3)}+\left[K_{ab}K^{ab}-({K_a}^a)^2+2\nabla_cJ^c\right] \\
J^c&=& n^c\nabla_a n^a-n^a\nabla_a n^c\\
NJ^c\nabla_c\phi&=&- h^{cd}D_c\phi D_d N+K \lp\dot{\phi}-N^a D_a\phi\rp\\
\sqrt{|g|}&=&N\sqrt{h}
\end{eqnarray}
The canonical variables of the theory are $(g_{ab},\phi)\equiv(N,N^a,h_{ab},\phi)$. The canonical momenta are defined by the following expressions 
\begin{eqnarray}
\Pi_N &=&\frac{\delta S}{\delta \dot{N}}=0 \ , \ \Pi_a=\frac{\delta S}{\delta \dot{N}^a}=0 \ , \\\
\Pi^{ab}&=&\frac{\delta S}{\delta \dot{h}_{ab}}=+\frac{\sqrt{h}}{2\kappa^2}\left[ \phi\lp K^{ab}-Kh^{ab}\rp-\frac{h^{ab}}{N}\lp\dot{\phi}-N^cD_c\phi\rp\right]\label{eq:momentumh} \ , \ \\
\pi_\phi&=&\frac{\delta S}{\delta\dot{\phi}}=\frac{\sqrt{h}}{2\kappa^2}\lp 2K+\frac{2\omega}{N\phi}\lp\dot{\phi}-N^cD_c\phi\rp\rp\label{eq:momentumphi}
\end{eqnarray}
Like in GR, we immediately see that the momenta conjugated to $N$ and $N^a$ are constrained to vanish. On the other hand, from the combination of $\Pi_h\equiv h_{ab}\Pi^{ab}$ and $\pi_\phi$, we find that 
\begin{equation} \label{eq:extra-constraint}
\Pi_h-\phi\pi_\phi=-\left(\frac{3+2w}{N}\right)\frac{\sqrt{h}}{2\kappa^2}\left(\dot\phi-N^cD_c\phi\right)
\end{equation}
is also constrained to vanish when $\omega=-3/2$, which is the case that interests us. It is now useful to rewrite the Lagrangian  density $\mathcal{L}$ using the definition for $\Pi^{ab}$ to eliminate the explicit dependence of $K_{ab}$ from it.  
The result is 
\begin{eqnarray}\label{eq:Lagrangian2}
\mathcal{L}&=&\frac{\sqrt{h}}{2\kappa^2}\left[N\left\{\phi R^{\lp3\rp}+\frac{1}{\phi}\frac{(2\kappa^2)^2}{h}\lp \Pi^{ab}\Pi_{ab}-\frac{\Pi^2_h}{2}\rp\right\}-\frac{N\omega}{\phi}D_c\phi D^c\phi+2D_c\phi D^cN-NV(\phi)\right. \nonumber\\
&+&\left.(3+2 w)\frac{\lp\dot{\phi}-N^cD_c\phi\rp^2}{2N\phi}\right] 
\end{eqnarray}
 Note that when $w=-3/2$, the last term in the above equation vanishes whereas it persists for $w\neq -3/2$ and can be expressed in terms of the momenta using (\ref{eq:extra-constraint}). A detailed discussion of both cases can be found in Ref.\refcite{OSA-2011}. Here we will just focus on the Palatini case, $w=-3/2$. To proceed with the construction of the Hamiltonian one must have in mind the above constraints and apply Dirac's algorithm\cite{PAMD} for constrained Hamiltonian systems. \\

Like in GR,  we have the primary constraints $C_N\equiv\Pi_N(t,x)=0$ and $C_a\equiv\Pi_a(t,x)=0$. Additionally, we have the constraint (\ref{eq:extra-constraint}). The Hamiltonian is constructed by introducing Lagrange multiplier fields $\lambda_N(t,x)$, $\lambda_a(t,x)$, and $\lambda_\phi$ for the primary constraints and performing the Legendre transform as usual with respect to the remaining {\it velocities}. The result is 
\begin{equation}
\bar{H}=\int d^3x \left[\lambda_N C_N+\lambda^a C_a+\lambda_\phi C_\phi+N^a\Ham_a+N\bar{\Ham}_N\right] \label{eq:H-Pal}
\end{equation}
where 
\begin{eqnarray}
C_N&=& \Pi_N \ , \ C_a=\Pi_a \ , \\
C_\phi&=&\Pi_h-\phi\pi_\phi \ , \\
\bar{\Ham}_N&=& \left(\frac{\sqrt{h}}{2\kappa^2}\right)\left[-\phi R^{(3)}+\frac{(2\kappa^2)^2}{h\phi}\left(\Pi^{ab}\Pi_{ab}-\frac{\Pi_h^2}{2}\right)+\frac{\omega}{\phi}D_c\phi D^c\phi+V(\phi)\right] \ , \label{eq:H_N_Pal} \\
\Ham_a&=&-2h_{ab}D_c\Pi^{bc}+\pi_\phi D_a\phi  \label{eq:H_Na}
\end{eqnarray}
For the dynamics to be consistent, the constraints must be preserved under evolution, which requires that $\dot{C}_N\equiv\{\bar{H},C_N\}=0$ and $\dot{C}_a\equiv\{\bar{H},C_a\}=0$, where the poisson bracket at time $t$ is defined as 
\begin{equation}
\{A(x),B(x')\}=\int d^3\sigma\left[\frac{\delta A(x)}{\delta \Pi^i(\sigma)}\frac{\delta B(x')}{\delta Q_i(\sigma)}-\frac{\delta B(x')}{\delta \Pi^i(\sigma)}\frac{\delta A(x)}{\delta Q_i(\sigma)}\right] \ ,
\end{equation}
where $\Pi^i$ and $Q_i$ generically represent the canonical variables. By direct evaluation, one finds that $\dot{C}_N=-\delta \bar{H}/\delta N=-\Ham_N$ and $\dot{C}_a=-\delta \bar{H}/\delta N^a=-\Ham_a$. We thus see that on consistency grounds we must impose the secondary constraints $\Ham_N=0$ and $\Ham_a=0$, which together with $C_\phi=0$ implies that the Hamiltonian $\bar{H}$ is constrained to vanish, like in GR. If matter is present, one must add the corresponding pieces $\delta H_{matt}/\delta N$ and $\delta H_{matt}/\delta N^a$ to these constraints, which leads to
\begin{equation}
\label{eq:constrN3/2}
-\phi R^{\lp 3\rp}+\frac{1}{\phi}\lp\tilde{\Pi}^{ab}\tilde{\Pi}_{ab}-\frac{\tilde{\Pi}^2_h}{2}\rp+\frac{\omega}{\phi}D_c\phi D^c\phi +2h^{cd}D_cD_d\phi+V(\phi)+\frac{1}{\alpha}\frac{\delta\mathcal{H}_{matt}}{\delta N}=0  
\end{equation}
\begin{equation}
-2D_c\tilde{\Pi}_a^c+\tilde{\pi}_\phi D_a\phi+\frac{1}{\alpha}\frac{\delta\mathcal{H}_{matt}}{\delta N^a}=0 \label{eq:constrNa3/2} \ ,
\end{equation}
where we have defined $\alpha\equiv h^{1/2}/(2\kappa^2)$ and used the tilde to denote the tensorial quantities $\tilde{\pi}_\phi=\pi_\phi/\alpha$ and $\tilde{\Pi}^{ab}=\Pi^{ab}/\alpha$. After some lengthy algebra, one finds the following evolution equations
\begin{eqnarray}\label{eq:dphi-Pal}
\dot{\phi}&=& N^aD_a\phi-\lambda_\phi \phi \\
\dot{\tilde{\pi}}_\phi&=& N\left[R^{(3)}+\frac{\tilde{\Pi}^{ab}\tilde{\Pi}_{ab}}{\phi^2}+\frac{w}{\phi^2}D_c\phi D^c\phi-\frac{dV}{d\phi}\right] \label{eq:dpi_phi-Pal}\\
&-& 2\Delta N+2wD_c\left(\frac{N}{\phi} D^c\phi\right)+N^aD_a\tilde{\pi}_\phi -\frac{\lambda_\phi\tilde{\pi}_\phi}{2} \nonumber\\
\dot{h}_{ab}&=&2D_{(a}N_{b)}+\frac{2N}{\phi}\left(\tilde{\Pi}_{ab}-\frac{h_{ab}}{2}\tilde{\Pi}_h\right)+\lambda_\phi h_{ab}\label{eq:dhab-Pal}\\
\dot{\tilde{\Pi}}^{ab}&=&-N\left[\phi \  ^{(3)}G^{ab}-\frac{w}{\phi}\left(D^a\phi D^b\phi-\frac{1}{2}h^{ab}D_c\phi D^c\phi\right)\right.\nonumber \\
&+&\left.\frac{2}{\phi}\left(\tilde{\Pi}^{ac}\tilde{\Pi}^b_c-\frac{h^{ab}}{4}\tilde{\Pi}^{mn}\tilde{\Pi}_{mn}\right)-\frac{\tilde{\Pi}_h}{2\phi}\left(3\tilde{\Pi}^{ab}-\frac{\tilde{\Pi}_h}{2}h^{ab}\right)\right] \nonumber\\
&+&N^cD_c \tilde{\Pi}^{ab}-\tilde{\Pi}^{ca}D_c N^{b}-\tilde{\Pi}^{cb}D_c N^{a}\nonumber \\
&+& D^aD^b(N\phi)-h^{ab}\Delta(N\phi)-2D^a N D^b \phi+h^{ab}D_cND^c\phi \nonumber \\
&-&\frac{NV}{2}h^{ab}-\frac{1}{\alpha}\frac{\delta H_{matt}}{\delta h_{ab}}-\frac{5}{2}\lambda_\phi \tilde{\Pi}^{ab} \ \label{eq:dPiab-Pal}.
\end{eqnarray}    
Using these evolution equations and the constraint (\ref{eq:constrN3/2}), one can verify that the evolution of $C_\phi$ leads to
\begin{equation}
\dot{C}_\phi=\{\bar{H},C_\phi\}=-2\alpha N V(\phi)+\alpha N\phi \frac{dV}{d\phi}-\frac{N}{2}\frac{\delta\mathcal{H}_{matt}}{\delta N}-h_{ab}\frac{\delta\mathcal{H}_{matt}}{\delta h_{ab}}\label{eq:Trace-constraint} 
\end{equation}
Since $\dot{C}_\phi$ must vanish, we must impose the secondary constraint
\begin{equation}
\phi \frac{dV}{d\phi}-2V(\phi)-\frac{1}{2\alpha}\frac{\delta\mathcal{H}_{matt}}{\delta N}-\frac{1}{N\alpha}h_{ab}\frac{\delta\mathcal{H}_{matt}}{\delta h_{ab}}=0 \label{eq:const3/2}
\end{equation}
Using the definitions $T_{ab}=-\frac{2}{\sqrt{-g}}\frac{\delta\mathcal{L}_{matt}}{\delta g^{ab}}$,  $g^{ab}=h^{ab}-\frac{1}{N^2}\lp t^a-N^a\rp\lp t^b-N^b\rp$, and the fact that $\frac{\delta\mathcal{L}_{matt}}{\delta N}=-\frac{\delta\mathcal{H}_{matt}}{\delta N}$ and $\frac{\delta\mathcal{L}_{matt}}{\delta h^{ab}}=-\frac{\delta\mathcal{H}_{matt}}{\delta h^{ab}}$, one can verify\cite{OSA-2011} that (\ref{eq:const3/2}) yields
\begin{equation}
\phi \frac{dV}{d\phi}-2V(\phi)=\kappa^2T \ . \label{eq:T-Ham}
\end{equation}
This equation reproduces the relation (\ref{eq:phi-BD}) when $w=-3/2$ and establishes an algebraic relation between the trace of the energy-momentum tensor of matter and the scalar field $\phi=\phi(T)$. \\

\subsubsection{Discussion}

From the derivations of above, we see that the dynamical variables in the Brans-Dicke case $w=-3/2$ are just $(h_{ab},\Pi^{ab})$ (plus the $(q_i,p^i)$ of the matter), because the evolution equations for $(\phi,\pi_\phi)$, as we saw above, can be combined to establish the secondary constraint (\ref{eq:T-Ham}). The lapse, $N$, and shift, $N^a$, manifest the diffeomorphism invariance of the theory and are not dynamical variables either. It is worth noting that the constraint (\ref{eq:constrN3/2}) involves up to second-order spatial derivatives of $h_{ab}$ (see the term $^{(3)}R$), but only first order time derivatives of it (contained in the momenta $\Pi^{ab}$). However, though that constraint contains spatial derivatives of $\phi=\phi(T)$ up to second order (see the term $D_aD_b\phi$), it does not contain any time derivative of $\phi(T)$ because the corresponding momentum $\pi_\phi$ is absent in that equation. Something analogous occurs in the vector constraint (\ref{eq:constrNa3/2}), where we can use the replacement $\pi_\phi=\Pi_h/\phi$ to show that no extra time derivatives of the matter appear in the constraints. This is a very important aspect, because it means that the highest order time derivative of the matter fields appearing in (\ref{eq:constrN3/2}) and (\ref{eq:constrNa3/2}) is the same as in GR and coincides with the highest order present in the energy-momentum tensor of the matter. The evolution equations also have this property. A glance at (\ref{eq:dphi-Pal}-\ref{eq:dPiab-Pal}) puts forward that the evolution equations for $\dot{\phi},\dot{h}_{ab}$, and $\dot{\Pi}^{ab}$ do not contain the momentum $\pi_\phi$, while in the equation for $\dot{\pi}_\phi$, one can reexpress the term $\lambda_\phi\tilde{\pi}_\phi$ using $\pi_\phi=\Pi_h/\phi$. Therefore, though one can find up to second-order spatial derivatives of $\phi(T)$, and hence of $T$, there is no trace of extra time derivatives acting on the matter fields. 
The existence of second-order spatial derivatives of $\phi(T)$ requires an extra degree of smoothness in the matter profiles, an aspect that is not necessary in GR. This extra degree of differentiability is a natural requirement if we attend to the $f(R)$ formulation of the $w=-3/2$ theory. Since the affine connection is compatible with a metric $t_{ab}$ which is conformally related with the space-time metric $g_{ab}$, the smoothness and differentiability of the conformal geometry is guaranteed if the conformal factor is differentiable up to second order (to yield a smooth field strength, Riemann tensor, of the affine connection). \\

  Let us now focus on the initial value problem. It is well-known that if in GR one specifies initial values for $N, N^a, h_{ab}$ and $\Pi^{ab}$ which are consistent with the constraint equations, the evolution equations uniquely determine $h_{ab}$ and $\Pi^{ab}$, while $N$ and $N^a$ remain undetermined, which expresses the existing gauge freedom of the theory. This guarantees that the intrinsic (coordinate-independent) geometry of space-time is determined uniquely\cite{MTW,Witten1962} by an initial choice of $h_{ab}$ and $\Pi^{ab}$ . The same is true for the scalar-tensor theories considered here, thus implying that the initial value problem is well-formulated\cite{OSA-2011} for all $w$. For the $w=-3/2$ case, the only difference with respect to GR is that one must specify an initial value for $\lambda_\phi$ taking into account its corresponding constraint equation to consistently establish the initial data. \\
Though the evolution equations presented here are not suitable to determine whether the initial value formulation is also well-posed, it is well-known that using different variables and representations of the evolution and constraint equations one can proof the well-posedness of GR and of generic Brans-Dicke theories with $w\neq -3/2$ in both Einstein and Jordan frames\cite{Salgado}. One can also make special choices for the lapse-shift pair and manipulate the corresponding $3+1$ equations of GR to show that the conjugate variables $h_{ab}$ and $\Pi^{ab}$ do satisfy a hyperbolic evolution system\cite{Choquet83}. One can thus exploit the resemblance between the constraint and evolution equations (\ref{eq:constrN3/2}),(\ref{eq:constrNa3/2}),(\ref{eq:T-Ham}) and (\ref{eq:dhab-Pal})-(\ref{eq:dPiab-Pal}) with those of GR to argue that the Cauchy problem is likely to be well-posed also for the Brans-Dicke case $w=-3/2$. 
Note first that in vacuum, $T_{\mu\nu}=0$ or $H_{matt}=0$, the constraint (\ref{eq:T-Ham}) implies that $\phi$ is a constant, $\phi_0$, which turns the constraints (\ref{eq:constrN3/2}) and (\ref{eq:constrNa3/2}) into
\begin{eqnarray}
\label{eq:constrNGRL}
&-&\phi_0 R^{\lp 3\rp}+\frac{1}{\phi_0}\lp\tilde{\Pi}^{ab}\tilde{\Pi}_{ab}-\frac{\tilde{\Pi}^2}{2}\rp+V(\phi_0)=0 \\
&-&2D_c\tilde{\Pi}_a^c=0 \label{eq:constrNaGRL} \ .
\end{eqnarray}
With a simple constant rescaling of the metric, these constraints are the same as those of GR with a cosmological constant. Setting for consistency the Lagrange multiplier $\lambda_\phi=0$, the evolution equations for $h_{ab}$ and $\tilde{\Pi}^{ab}$ also recover the same form as those of GR with a cosmological constant. We can thus conclude that the Cauchy problem in vacuum is well-posed. \\
When matter is present, one should add to the above equations those corresponding to the matter fields. The strategy now would be to interpret the $\phi$-dependent terms, which are functions of the trace $T$, as part of a new (or modified) matter Hamiltonian. This way, the constraint and evolution equations maintain a structure that closely resembles that of GR except by some non-constant factors $\phi(T)$ that multiply or divide objects like $^{(3)}R$ and $\tilde{\Pi}^{ab}\tilde{\Pi}_{ab}$. If the matter fields satisfy the spatial differentiability requirements imposed by the constraint equations, the absence of higher-order time derivatives of the matter fields in the constraint and evolution equations suggests that the time evolution will be as well-posed as in GR. This, in fact, has been explicitly shown for a perfect fluid using the Einstein frame representation of the evolution equations\cite{Capo-Vignolo}. Obviously, since in general the well-posedness of the GR equations depends on the particular matter sources considered, the modification of the source terms induced by the existence of $\phi(T)$-dependent terms requires a model by model analysis. Therefore, though one cannot conclude that the Cauchy problem is well-posed for an arbitrary $f(R)$ Palatini Lagrangian, we find no reasons to suspect that it is ill-posed in general.\\ 

To close this section, we comment on recent literature that criticizes the viability of {\it all} Palatini $f(R)$ theories based on a seemingly ill-formulation of the Cauchy problem. In Ref.\refcite{Far07} it was claimed that the disappearance of the d'Alambertian $\Box \phi$ from (\ref{eq:phi-BD}) for the value $w=-3/2$ implies that the non-dynamical field $\phi$ can be arbitrarily assigned on a region or on the entire spacetime, provided its gradient satisfies a degenerate equation [Eq. $(4.5)$ in that paper], which reduces to a constraint. This fact, it was stated, would make impossible to eliminate the term $\Box \phi$ from the evolution equations unless $\Box\phi=0$. This was interpreted as a {\it no-go theorem} for Palatini $f(R)$ gravity, which would  have an ill-formulated Cauchy problem even in vacuum. This interpretation is conceptually wrong (see also Ref.\refcite{Faraoni:2008bu} in this respect) because the scalar field in the $w=-3/2$ is just a given algebraic function of the trace $T$ and, therefore, is clearly specified by the local matter content\footnote{The fact that the amplitude of the scalar field when $w=-3/2$ is determined algebraically by the local matter sources also implies that the effective Newton's constant $G_{eff}=G/\phi$ is only subject to local variations of the energy-momentum densities. In this sense, though $G_{eff}$ does change over cosmic timescales due to the expansion of the universe, it is not subject to the same type of time evolution that affects the $w\neq -3/2$ Brans-Dicke theories and other dynamical scalar-tensor theories, which contrasts with the interpretation of Ref.\refcite{Li:2009zz}. }. Moreover, one should note that  Eq.$(4.5)$ of Ref.\refcite{Far07} is not correct. That equation should recover the well-known relation $2V-\phi V'=\kappa^2T$ that establishes  the algebraic relation between $\phi$ and $T$ [the secondary constraint (\ref{eq:T-Ham})]. Using Eqs. $(3.4)$ and $(3.5)$ of Ref.\refcite{Far07}, it is easy to check that the associated Eq. $(3.10)$ does recover our equation (\ref{eq:T-Ham}) in the Brans-Dicke case $w=-3/2$ (even though this is not the result obtained in Ref.\refcite{Far07}). This indicates that the first claims against the well-posedness of the Cauchy problem for Palatini $f(R)$ theories stemmed from a misleading analysis of erroneous equations.\\
The strong conclusions of Ref.\refcite{Far07} were a bit relaxed in Ref.\refcite{FarSot08} (see in this sense Refs.\refcite{Capo09,Far09}), where it was admitted that the Cauchy problem should be well-posed in vacuum and with radiation fields (for which $T=0$ and $\phi=$constant). In fact, in Ref.\refcite{FarSot08} it was correctly noticed that in the $w=-3/2$ case the field $\phi$ could be algebraically solved in terms of $T$ (though their Eq. $(219)$ is the same as Eq.$(4.5)$ of Ref.\refcite{Far07}). It was then argued that the existence of terms of the form $\Box \phi(T)$, which imply contributions of the form $\Box T$, would cause problems for the Cauchy problem. Though such terms and the possible existence of higher-order derivatives of the matter fields are certainly a reason for concern, it was prematurely concluded that the Cauchy problem for Palatini $f(R)$ theories was likely to be neither well-formulated nor well-posed unless the trace $T$ were constant. These conclusions contrast with the findings of Ref.\refcite{OSA-2011} presented here, which show that the evolution equations do not introduce higher-order time derivatives of the matter fields, which guarantees that the initial value problem is as well formulated\cite{MTW,Witten1962} as in GR.

\section{Nonsingular bouncing cosmologies \label{sec:QG}}

We have seen that cosmological observations and local experiments strongly constrain the form of the $f(R)$ gravity Lagrangian at low curvatures
(see Refs.\refcite{Koivisto-2006,AEMM-2006,Li:2006ag,Carvalho:2008am,Santos:2008qp,Pires-2010,Baghram,Li-Chu-2006,Koi-Kurki-2006,ULT-2007,FayTT-2006,TUT-2008} for cosmological constraints and Refs.\refcite{Olmo2005,Olmo2007b,Flanagan-2004a,Olmo-2008a,LMS-2008,LMS-2009} for local experiments). Though many $f(R)$ models have the ability to produce late-time cosmic acceleration and fit well the background expansion history, they are not in quantitative agreement with the structure and evolution of cosmic inhomogeneities. Additionally, we have seen that the fact that matter is concentrated in discrete structures like atoms causes the modified dynamics to manifest also in laboratory experiments, which confirms earlier suspicions on the viability of such models according to their corresponding Newtonian and post-Newtonian properties. 
This is a very disturbing aspect of the models with infrared corrections, which demands the consideration of a microscopic description of the sources and prevents the use of macroscopic, averaged representations of the matter. A careful analysis of this point put forward the existence of non-perturbative effects induced by the Palatini dynamics in a number of contexts\cite{Flanagan-2004a,Olmo2007b,IKPP,Sot08a,Sot08b,Olmo-2008a}. In this sense, it is worth noting that even though the ground state of Hydrogen can be studied using standard perturbative methods, the first and higher excited states do manifest non-perturbative properties\cite{Olmo-2008a}.  
Despite the fact that the modified dynamics is strongly suppressed in regions of high density, non-perturbative effects arise near the zeros of the atomic wavefunctions, where the matter density crosses the characteristic low-density scale of the theory and the gradients of the matter distribution become very important for the dynamics [see eq.(\ref{eq:Gab-f})]. Though this certainly is an undesired property of infrared-corrected models, it could become a very useful tool for models with corrections at high curvatures. Can we construct singularity-free cosmological models that recover GR at low curvatures using the non-perturbative properties of Palatini theories? As we will see, ultraviolet-corrected Palatini models turn out to be very efficient at removing the big bang cosmic singularity in various situations of interest. In this section we will thus review recent efforts carried out to better understand the properties of Palatini theories in the early universe. \\

\subsection{Non-singular $f(R)$ cosmologies}

\begin{figure}[ht]
\begin{center}
{\psfig{file=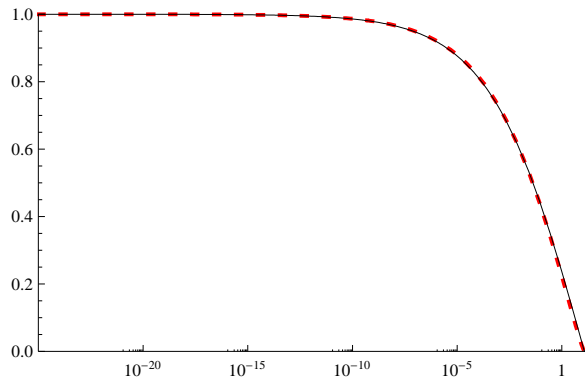,width=0.6\textwidth}} 
\vspace*{8pt}
\caption{Vertical axis: {$df/dR$} \ ; \ Horizontal axis: {$R/R_c$}. Comparison of the numerical solution with the interpolating function (\ref{eq:f-guess}). The dashed line represents the numerical curve.\label{LQC-plot}}
\end{center}
\end{figure}

Growing interest in the dynamics of the early-universe in Palatini theories has arisen, in part, from the observation that the effective equations of loop quantum cosmology\cite{LQC} (LQC), a Hamiltonian approach to quantum gravity based on the quantization techniques of loop quantum gravity, could be exactly reproduced by a Palatini $f(R)$ Lagrangian\cite{OS-2009}. In LQC, non-perturbative quantum gravity effects lead to the resolution of the big bang singularity by a quantum bounce without introducing any new degrees of freedom. Though fundamentally discrete, the theory admits a continuum description in terms of an effective Hamiltonian that in the case of a homogeneous and isotropic universe filled with a massless scalar field leads to the following modified Friedmann equation
\begin{equation}\label{eq:LQC}
3H^2={8\pi G}\rho\left(1-\frac{\rho}{\rho_{crit}}\right) \ ,
\end{equation}  
where $\rho_{crit}\approx 0.41\rho_{Planck}$. At low densities, $\rho/\rho_{crit}\ll 1$, the background dynamics is the same as in GR, whereas at densities of order $\rho_{crit}$ the non-linear new matter contribution forces the vanishing of $H^2$ and hence a cosmic bounce. This singularity avoidance seems to be a generic feature of loop-quantized universes\cite{Param09}. \\
Palatini $f(R)$ theories share with LQC the fact that the modified dynamics that they produce is not due to the existence of new dynamical degrees of freedom but rather to non-linear effects produced by the matter sources, which contrasts with other approaches to quantum gravity and to modified gravity. This similarity makes it tempting to put into correspondence Eq.(\ref{eq:LQC}) with the corresponding $f(R)$ equation
\begin{equation}
3H^2=\frac{f_R\left(\kappa^2\rho+({R}f_R-f)/2\right)}{\left(f_R-\frac{12\kappa^2\rho f_{RR}}{2 ({R}f_{RR}-f_R)}\right)^2} \ .
\end{equation}
Taking into account the trace equation (\ref{eq:trace-f}), which for a massless scalar becomes $R f_R -2f=2\kappa^2\rho$ and implies that $\rho=\rho({R})$, one finds that a Palatini $f(R)$ theory able to reproduce the LQC dynamics (\ref{eq:LQC}) must satisfy the differential equation
\begin{equation}
f_{RR}=-f_R\left(\frac{A f_R -B}{2({R}f_R-3f)A+{R}B}\right)
\end{equation}
where $A=\sqrt{2({R}f_R-2f)(2{R}_c-[Rf_R-2f])}$, $B=2\sqrt{{R}_cf_R(2{R} f_R-3f)}$, and ${R}_c\equiv \kappa^2\rho_c$. If one imposes the boundary condition $\lim_{R\to 0} f_R\to 1$ at low curvatures, and $\ddot{a}_{LQC}=\ddot{a}_{Pal}$ (where $\ddot{a}$ represents the acceleration of the expansion factor) at $\rho=\rho_c$, the solution to this equation is unique.  The solution was found numerically\cite{OS-2009}, though the following function can be regarded as a good approximation to the LQC dynamics from the GR regime to the non-perturbative bouncing region (see Fig.\ref{LQC-plot})
\begin{equation}\label{eq:f-guess}
\frac{df}{dR}=- \tanh \left(\frac{5}{103}\ln\left[\left(\frac{R}{12\mathcal{R}_c}\right)^2\right]\right)
\end{equation} 
It should be noted that different attempts to find effective actions for the LQC equations have also been considered but either failed\cite{BR-2009}
or are limited to the low-energy, perturbative regime\cite{Sot-LQC}.\\

\begin{figure}[ht]
\begin{center}
{\psfig{file=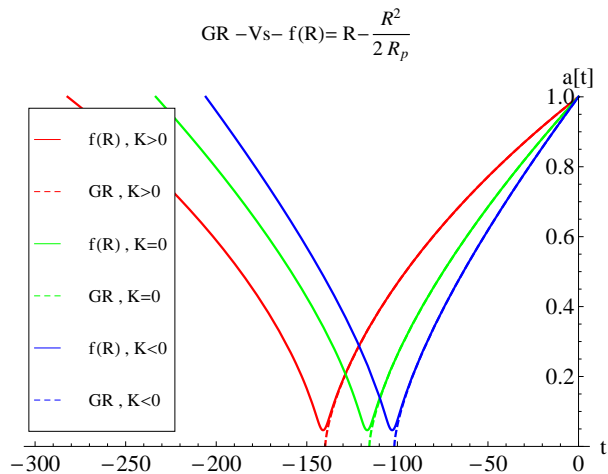,width=0.8\textwidth}} 
\caption{Time evolution of the expansion factor for the model $f(R)=R-R^2/2R_P$ and $w=0$ for $K>0$, $K=0$, and $K<0$ (solid curves from left to right). From left to right, we see that the universe is initially contracting, reaches a minimum, and then bounces into an expanding phase. The dashed lines, which are only discernible near the bounces, represent the expanding solutions of GR, which begin with a big bang singularity ($a(t)=0$) and quickly tend to the nonsingular solutions.  \label{fig:fR_K_a}}
\end{center}
\end{figure}

Though the function (\ref{eq:f-guess}) implies that the LQC Lagrangian is an infinite series, which is a manifestation of the non-local properties of the quantum geometry, the fact is that one can find non-singular cosmologies of the $f(R)$ type with a finite number of terms. In fact, a simple quadratic Lagrangian of the form $f(R)=R+R^2/R_P$ does exhibit non-singular solutions\cite{SS-1990} for certain equations of state\cite{BOSA-2009,MyTalks} depending on the sign of $R_P$. To be precise, if $R_P > 0$ the bounce occurs for sources with $w=P/\rho> 1/3$. If $R_P < 0$, then the bouncing condition is satisfied by $w < 1/3$ (see Fig.\ref{fig:fR_K_a}). This can be easily understood by having a look at the expression for the Hubble function in a universe filled with radiation plus a fluid with generic equation of state $w$ and density $\rho$
\begin{equation}\label{eq:Hubble-iso}
H^2=\frac{1}{6f_R}\frac{\left[f+(1+3w)\kappa^2\rho+2\kappa^2\rho_{rad}-\frac{6K f_R}{a^2}\right]}{\left[1+\frac{3}{2}\Delta_1\right]^2} 
\end{equation}
where ${\Delta}_1=-(1+w)\rho\partial_\rho f_R/f_R=(1+w)(1-3w)\kappa^2\rho  f_{RR}/(f_R(Rf_{RR}-f_R))$. Due to the structure of $\Delta_1$, one can check that $H^2$ vanishes when $f_R\to 0$. A more careful analysis\cite{BO-2010} shows that $f_R\to 0$ is the only possible way to obtain a bounce with a Palatini $f(R)$ theory that recovers GR at low curvatures if $w$ is constant. In the case of $f(R)=R+R^2/R_P$, it is easy to see that $f_R=0$ has a solution if $1+2R_{Bounce}/R_P=0$ is satisfied for $\rho_{Bounce}>0$, where $R_{Bounce}=(1-3w)\kappa^2\rho_{Bounce}$, which leads to the cases mentioned above. It is worth noting, see Fig.\ref{fig:fR_K_a}, that the expanding branch of the non-singular solution rapidly evolves into the solution corresponding to GR. The departure from the GR solution is only apparent very near the bounce, which is a manifestation of the non-perturbative nature of the solution. Note also that in GR there is a solution that represents a contracting branch that ends at the singularity  where the expanding branch begins (this solution is just the time reflection of the expanding branch). The Palatini model $f(R)=R-R^2/2R_P$ represented here simply allows for a smooth transition from the initially contracting branch to the expanding one. 

Besides avoiding the development of curvature singularities, bouncing cosmologies can solve the horizon problem\cite{Novello-2008}, which makes them interesting as a substitute for inflation. To be regarded as a serious candidate to explain the phenomenology of the early universe, these theories should provide a consistent evolution of perturbations across the bounce, which should also be compatible with the observed nearly scale invariant spectrum of primordial perturbations. Investigations in this direction have found\cite{Koi-2010} that $f(R)$ models that develop a bounce when the condition $f_R=0$ is met turn out to exhibit singular behavior of inhomogeneous perturbations in a flat, dust-filled universe. However, since some terms in the perturbation equations blow up as $f_R\to 0$, their backreaction renders the perturbative system invalid and, therefore, one cannot say if there is a true singularity or not.\\

Further insight on the robustness of the bounce under perturbations  was obtained\cite{BO-2010} studying the properties of $f(R)$ theories in anisotropic spacetimes of Bianchi-I type
\begin{equation}
ds^2=-dt^2+\sum_{i=1}^3 a_i^2(t)(dx^i)^2 \ .
\end{equation} 
If one considers these space-times under the dynamics of Palatini theories with a generic perfect fluid, one can derive a number of useful analytical expressions. In particular, one finds that the expansion $\theta=\sum_i H_i$ and the shear $\sigma^2=\sum_i\left(H_i-\frac{\theta}{3}\right)^2$ (a measure of the degree of anisotropy) are given by 
\begin{equation}\label{eq:Hubble-f(R)}
\frac{\theta^2}{3}\left(1+\frac{3}{2}\Delta_1\right)^2=\frac{f+\kappa^2(\rho+3P)}{2f_R}+\frac{\sigma^2}{2}
\end{equation}
\begin{equation}\label{eq:shear-f(R)}
\sigma^2=\frac{\rho^{\frac{2}{1+w}}}{f_R^2}\frac{(C_{12}^2+C_{23}^2+C_{31}^2)}{3} \ ,
\end{equation}
where the constants $C_{ij}=-C_{ji}$ set the amount and distribution of anisotropy and satisfy the constraint $C_{12}+C_{23}+C_{31}=0$. In the isotropic case, $C_{ij}=0$, one has $\sigma^2=0$ and $\theta^2=9H^2$, with $H^2$ given by Eq.(\ref{eq:Hubble-iso}). Now, 
since homogeneous and isotropic bouncing universes require the condition $f_R=0$ at the bounce, a glance at (\ref{eq:shear-f(R)}) 
indicates that the shear diverges as $\sim 1/f_R^2$. This shows that, regardless of how small the anisotropies are initially, any isotropic $f(R)$ bouncing model will develop divergences when anisotropies are present. It is worth noting that even though $\sigma^2$ diverges at $f_R=0$, the expansion and its time derivative\cite{BO-2010} are smooth and finite functions at that point if the density and curvature are finite. However, 
one can check by direct calculation that the Kretschman scalar $R_{\mu\nu\sigma\rho}R^{\mu\nu\sigma\rho}=4(\sum_i(\dot H_i+H_i^2)^2+H_1^2H_2^2+H_1^2H_3^2+H_2^2H_3^2)$ diverges at least as $\sim 1/f_R^4$, which is a clear geometrical pathology and signals the presence of a physical singularity. The problems when $f_R$ vanishes  seem to be generic in anisotropic models of modified theories of gravity\cite{FF-Saa09}.\\

\subsection{Nonsingular cosmologies beyond $f(R)$ \label{sec:beyond}}

The previous section provides reasons to believe that Palatini $f(R)$ models are not able to produce a fully satisfactory and singularity-free alternative to GR in idealized universes filled with a single perfect fluid with constant equation of state\footnote{The consideration of several fluids, fluids with varying equation of state, or fluids with anisotropic stresses, see for instance\cite{Koivisto07}, could affect the dynamics providing new bouncing mechanisms and preventing the extension of this conclusion to such more realistic cases.}. Though the homogeneous and isotropic case greatly improves the situation with respect to GR, the existence of divergences when anisotropies and inhomogeneities are present spoil the hopes deposited on this kind of Lagrangians. To the light of these results, new Palatini theories were explored\cite{BO-2010} to determine if the introduction of new elements in the gravitational action could avoid the problems that appear in the $f(R)$ models. This led to the study of isotropic and anisotropic cosmologies of some simple generalization of the $f(R)$ family in which the Lagrangian takes the form $f(R,Q)$, with $Q=R_{\mu\nu}R^{\mu\nu}$. Using the particular Lagrangian 
\begin{equation}\label{eq:f(R,Q)}
f(R,R_{\mu\nu}R^{\mu\nu})=R+a\frac{R^2}{R_P}+\frac{R_{\mu\nu}R^{\mu\nu}}{R_P} \ ,
\end{equation}
where $R_P\sim l_P^{-2}$ is the Planck curvature, it was found that completely regular bouncing solutions exist for both isotropic and anisotropic homogeneous cosmologies filled with a perfect fluid. In particular, one finds that for $a<0$ the interval $0\leq w\leq 1/3$ is always included in the family of bouncing solutions, which contains the dust and radiation cases. For $a\geq 0$, the fluids yielding a non-singular evolution are restricted to $w>\frac{a}{2+3a}$, which implies that the radiation case $w=1/3$ is always nonsingular. For a detailed discussion and classification of the non-singular solutions depending on the value of the parameter $a$ and the equation of state $w$, see Ref.\refcite{BO-2010}. \\

\begin{figure}[ht]
\begin{center}
{\psfig{file=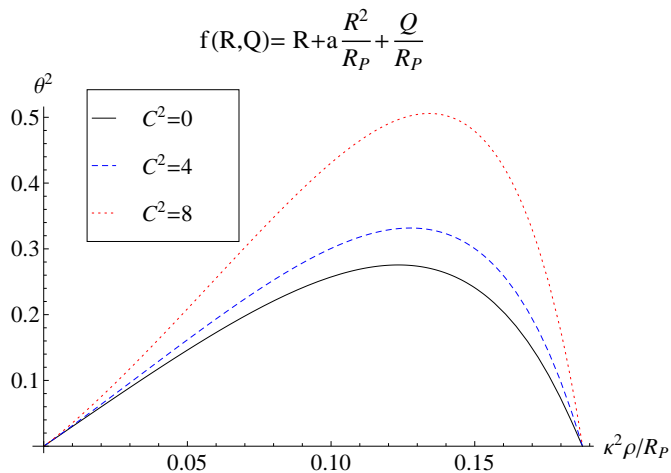,width=0.8\textwidth}} 
\caption{Evolution of the expansion as a function of $\kappa^2\rho/R_P$ in radiation universes with low anisotropy, which is controlled by the combination $C^2=C_{12}^2+C_{23}^2+C_{31}^2$. The case with $C^2=0$ corresponds to the isotropic flat case, $\theta^2=9H^2$.  \label{fig:ExpanRad}}
\end{center}
\end{figure}

The field equations that follow from the Lagrangian (\ref{eq:f(R,Q)}) when $R_{\mu\nu}$ is assumed symmetric\footnote{See Ref.\refcite{Vitagliano:2010pq} for the case when this condition is relaxed.} were derived in Ref.\refcite{OSAT} (see also Refs.\refcite{LMS-2008,OSAT-MG}) and take the form 
\begin{eqnarray}\label{eq:met-var}
f_R R_{\mu\nu}-\frac{f}{2}g_{\mu\nu}+2f_QR_{\mu\alpha}{R^\alpha}_\nu &=& \kappa^2 T_{\mu\nu} \ , \\
\nabla_{\beta}\left[\sqrt{-g}\left(f_R g^{\mu\nu}+2f_Q R^{\mu\nu}\right)\right]&=&0 \label{eq:con-var}
\end{eqnarray}
where $f_R\equiv \partial_R f$ and $f_Q\equiv \partial_Q f$. The connection equation (\ref{eq:con-var}) can be solved in general introducing an auxiliary metric $h_{\alpha\beta}$ such that (\ref{eq:con-var}) takes the form $\nabla_{\beta}\left[\sqrt{-h} h^{\mu\nu}\right]=0$, which implies that $\Gamma^{\rho}_{\mu\lambda}$ can be written as the Levi-Civita connection of $h_{\mu\nu}$. When the matter sources are represented by a perfect fluid, $T_{\mu\nu}=(\rho+P)u_\mu u_\nu+P g_{\mu\nu} $, one can show that $h_{\mu\nu}$ and $h^{\mu\nu}$ are given by\cite{OSAT}
\begin{eqnarray}\label{eq:met-disformal}
h_{\mu\nu}&=&\Omega\left( g_{\mu\nu}-\frac{\Lambda_2}{\Lambda_1-\Lambda_2} u_\mu u_\nu \right)\\
h^{\mu\nu}&=&\frac{1}{\Omega}\left( g^{\mu\nu}+\frac{\Lambda_2}{\Lambda_1} u^\mu u^\nu \right)
\end{eqnarray}
where 
\begin{eqnarray}
\Omega&=&\left[\Lambda_1(\Lambda_1-\Lambda_2)\right]^{1/2} \ , \ \lambda=\sqrt{\kappa^2 P+\frac{f}{2}+\frac{f_R^2}{8f_Q}} \\
\Lambda_1&=& \sqrt{2f_Q}\lambda+\frac{f_R}{2}  \ , \ \Lambda_2= \sqrt{2f_Q}\left[\lambda\pm\sqrt{\lambda^2-\kappa^2(\rho+P)}\right] 
\end{eqnarray}
It is worth noting that (\ref{eq:met-disformal}) implies a disformal relation between the metrics $g_{\mu\nu}$ and $h_{\mu\nu}$. A relation of this form between two metrics naturally arises in Bekenstein's relativistic theory\cite{bekenstein} of MOND and in previous versions of it. In the MOND theory, the vector $u_\mu$ is an independent dynamical vector field and the functions in front of it and in front of $g_{\mu\nu}$ depend on another 
dynamical scalar field. In the theory described here, on the contrary, the metric tensor is the only dynamical field of the gravitational sector.
Note also that a Palatini-like version of MOND has been recently proposed by Milgrom\cite{Milgrom:2009ee}.\\
In terms of $h_{\mu\nu}$ and the above definitions, the metric field equation (\ref{eq:met-var}) takes the following form 
\begin{equation}\label{eq:Rmn-h}
R_{\mu\nu}(h)=\frac{1}{\Lambda_1}\left[\frac{\left(f+2\kappa^2P\right)}{2\Omega}h_{\mu\nu}+\frac{\Lambda_1\kappa^2(\rho+P)}{\Lambda_1-\Lambda_2}u_{\mu}u_{\nu}\right] \ .
\end{equation}
In this expression, the functions $f, \Lambda_1$, and $\Lambda_2$ are functions of the density $\rho$ and pressure $P$. In particular, for our quadratic model one finds that $R=\kappa^2(\rho-3P)$, like in GR, and $Q=Q(\rho,P)$ is given by 
\begin{equation}\label{eq:Q}
\frac{Q}{2R_P}=-\left(\kappa^2P+\frac{\tilde f}{2}+\frac{R_P}{8}\tilde f_R^2\right)+\frac{R_P}{32}\left[3\left(\frac{ R}{R_P}+\tilde f_R\right)-\sqrt{\left(\frac{R}{R_P}+\tilde f_R\right)^2-\frac{ 4 \kappa^2(\rho+P)}{R_P} }\right]^2 \ ,
\end{equation}
where $\tilde f=R+aR^2/R_P$, and the minus sign in front of the square root has been chosen to recover the correct limit at low curvatures. In a universe filled with radiation, for which $R=0$, the function $Q$ boils down to\cite{BO-2010}  
\begin{equation}
Q= \frac{3R_P^2}{8}\left[1-\frac{8\kappa^2\rho}{3R_P}-\sqrt{1-\frac{16\kappa^2\rho}{3R_P}}\right] \label{eq:Q-rad} \ .
\end{equation}
This expression recovers the GR value at low curvatures, $Q\approx 4(\kappa^2\rho)^2/3+32(\kappa^2\rho)^3/9R_P+\ldots$ but reaches a maximum $Q_{max}=3R_P^2/16$ at $\kappa^2\rho_{max}=3R_P/16$, where the squared root of (\ref{eq:Q}) vanishes. At $\rho_{max}$ the shear also takes its maximum allowed value, namely, $\sigma^2_{max}=\sqrt{3/16}R_P^{3/2}(C_{12}^2+C_{23}^2+C_{31}^2)$, which is always finite, and the expansion vanishes producing a cosmic bounce regardless of the amount of anisotropy (see Fig.\ref{fig:ExpanRad}). The model (\ref{eq:f(R,Q)}), therefore, avoids the well-known problems of anisotropic universes in GR\cite{ekpyrotic}, where anisotropies grow faster than the energy density during the contraction phase leading to a singularity that can only be avoided by sources with $w>1$.

The evolution of inhomogeneities in the quadratic model discussed here was considered in Ref.\refcite{LBM07}, though the approximations used there to solve for the connection equation did not allow to see the existence of bouncing solutions. For this reason, in this case one cannot make any statement regarding the evolution of inhomogeneities across the bounce. The cosmology of $f(R)$ and $f(R_{\mu\nu}R^{\mu\nu})$ theories was also considered in some detail in Ref.\refcite{ABF04}. The possibility of having a standard cosmological evolution in $f(R,Q)$ models with a large cosmological constant has been considered recently\cite{Bauer:2010bu}. \\
It should be noted that the choice of a symmetric Ricci tensor in the analysis of $f(R,Q)$ bouncing cosmologies presented above is not arbitrary. As shown in Ref.\refcite{Vitagliano:2010pq}, the antisymmetric part of the Ricci tensor introduces new dynamical degrees of freedom in the form of a massive vector field (see also Ref.\refcite{Vollick:2006uq} for a related result). If one looks for a framework suitable for the description of the effective dynamics of LQC (including anisotropies) and, more generally, of other theories of quantum geometry not involving new degrees of freedom, it seems natural to impose constraints on the spectrum of possible Lagrangians to avoid new propagating fields. In this sense, we note that the $f(R,Q)$ theories discussed here are able to reproduce\cite{Olmo-2011} other aspects of the expected phenomenology of quantum gravity at the Planck scale. In particular, without imposing any a priori phenomenological structure, the quadratic Palatini model (\ref{eq:f(R,Q)}) predicts an energy-density dependence of the metric components that closely matches the structure conjectured in models of Doubly (or Deformed) Special Relativity\cite{DSR} and Rainbow Gravity\cite{RG}. This confirms that Palatini theories represent a new and powerful framework to address different aspects of quantum gravity phenomenology.

\section{Summary and Conclusions}

From the number of works that have been discussed in this review, it seems fair to say that the Palatini approach to modified gravity has experienced a recent period of accelerated expansion motivated by theoretical and observational advances in cosmology. The possibility of explaining the cosmic speedup problem in geometrical terms boosted the interest in all sorts of modified theories of gravity with special emphasis in the $f(R)$ family. Palatini $f(R)$ theories appeared at first as an exotic alternative to the more familiar metric formulation of those theories. They had the advantage of naturally producing an effective cosmological constant\cite{Vollick-2003}, of avoiding certain dynamical instabilities present in their metric formulation\cite{f(R),SotPLB}, and of yielding second-order evolution equations. However, the first models chosen to attack the cosmic acceleration problem (see Sec.\ref{sec:speedup}) had the undesired feature of requiring a microscopic description of the matter sources (see Sec.\ref{sec:atoms}). The analysis of the dynamics of those models in the microscopic world put forward the existence of non-perturbative effects which seemed to be in clear conflict with our understanding of the physics at small scales. To overcome the technical difficulties posed by this situation, different directions were followed to test the viability of various families of $f(R)$ models. This motivated the analysis of the weak field limit\cite{Olmo2005,Meng-Wang-Newton,DomBarraco,AFRT-2005}, the subtleties in the description of averaged distributions of matter\cite{Flanagan-2004a,Olmo-2008a,LMS-2008,LMS-2009}, the stability and structure of stellar objects\cite{Kainu07a,Sot08a,Kainu07b,Sot08b,Olmo-2008b,Bar08,Reij-2009}, the Cauchy problem\cite{Far07,OSA-2011,Capo-Vignolo}, and other issues that complemented the continuous investigation of the cosmological dynamics of these theories. All these different approaches have raised interesting and healthy debates that have shed light on the many scenarios in which the gravitational dynamics of Palatini theories may have an influence. From those debates it follows that the background expansion history of viable $f(R)$ models is currently (statistically) indistinguishable from that of the standard $\Lambda$CDM model (GR with cold dark matter and a cosmological constant), that laboratory and solar system tests can efficiently put constraints on model parameters, that stellar structure can also be used to set some constraints on $f(R)$ models, that the Cauchy problem is well-formulated and well-posed in many situations of interest, and that Palatini theories are a powerful new tool to address different aspects of quantum gravity phenomenology. \\

The observation that Palatini theories can be used to describe the effective geometry of space-times with a discrete quantum structure\cite{OS-2009} provides solid reasons to explore the properties of $f(R)$ and more general theories in the early universe and in scenarios involving strong gravitational fields and very high energy densities. In this sense, we note that simple extensions of the $f(R)$ family that include Ricci squared terms\cite{OSAT,BO-2010,Vitagliano:2010pq,OSAT-MG,LBM07} present a much richer phenomenology than $f(R)$ models. On the other hand, the fact that the dark matter problem in galaxies can be addressed from a new class\cite{Milgrom:2009ee} of Palatini theories suggests that new approaches to the dark matter and dark energy problems beyond the $f(R)$ family are possible. The exploration of the field equations, cosmology, black hole formation, stellar structure, galactic dynamics, \ldots  of new and more general Palatini theories will surely yield interesting new results with potential applications to quantum gravity, the late-time cosmology, and astrophysics. We hope that this review helps active researchers in this field and encourages newcomers to continue the exploration of the Palatini approach to modified gravity to address and solve some of the important problems that cosmology faces nowadays.\\

{\bf Acknowledgements.} Work supported by the Spanish grant FIS2008-06078-C03-02, and the Consolider Programme CPAN (CSD2007-00042). 


\end{document}